\documentclass[10pt]{article}
\usepackage{amsmath,amssymb,subfigure}
\usepackage[dvips]{epsfig}
   
\newtheorem{lem}{Lemma}[section]

\newtheorem{rem}[lem]{Remark}

\newtheorem{prop}[lem]{Proposition}
\newtheorem{theo}[lem]{Theorem}
\newcommand{\proof}{{\bf Proof}:\quad}
\newcommand{\nc}{\newcommand}
\newcommand{\rnc}{\renewcommand}
\nc{\beqn}{\begin{eqnarray*}}
\nc{\eeqn}{\end{eqnarray*}}
\nc{\be}{\begin{equation}}
\nc{\ee}{\end{equation}}
\nc{\beqa}{\begin{eqnarray}}
\nc{\eeqa}{\end{eqnarray}}
\rnc{\a}{\alpha}
\rnc{\d}{\delta}
\nc{\ga}{\gamma}
\nc{\lb}{\lambda}
\nc{\f}{\phi}
\nc{\p}{\psi}
\nc{\e}{\epsilon}
\rnc{\c}{\chi}
\nc{\sg}{\sigma}
\rnc{\t}{\theta}
\nc{\om}{\omega}
\rnc{\P}{\Psi}
\nc{\G}{\Gamma}

\nc{\ra}{\rightarrow}
\nc{\Ra}{\Rightarrow}
\nc{\LRa}{\Leftrightarrow}
\nc{\lra}{\leftrightarrow}
\nc{\ot}{\otimes}

\nc{\mat}[4]{\left(\begin{array}{cc}#1&#2\\#3&#4\end{array}\right)}

\nc{\ca}{\mathcal{A}}
\nc{\cf}{\mathcal{F}}
\nc{\cn}{\mathcal{N}}
\nc{\cs}{\mathcal{S}}
\nc{\Gt}{\tilde{\Gamma}}
\newcommand{\dozr}{\omega^g_z(r)}

\nc{\csh}{\hat{\cs}_g}
\nc{\dorz}{\omega^g_r(z)}
\nc{\dhorz}{\hat{\omega}^g_r(z)}
\newcommand{\doq}{\omega_{qq_0}}
\newcommand{\dop}{\omega_{pp_0}}

\newcommand{\rres}{\operatornamewithlimits{Res}}
\newcommand{\res[1]}{\operatornamewithlimits{Res}_{#1}}

\newcommand{\resi}{\operatornamewithlimits{Res}_{z=\infty}}

\newcommand{\vu}{{\vec{u}}}
\nc{\VU}{{\vec{U}}}

\newcommand{\Rn}{{\rm I\!R}}
\newcommand{\Nn}{{\rm I\!N}}
\newcommand{\Pn}{{\rm I\!P}}
\begin{document}

\title{From the solution of the Tsarev system to the solution of the Whitham equations}
\author{Tamara Grava\thanks{Present address: Department of Mathematics, Imperial
College, London SW7 2BZ UK, e-mail: t.grava@ic.ac.uk.}\\Department
of Mathematics, University of Maryland\\
College Park 20742-4015\\ USA}
\maketitle
\vskip 0.7cm
\begin{abstract}
\noindent
We study the Cauchy problem for the Whitham modulation equations  for
monotone increasing smooth initial data.
The Whitham equations  are a collection of one-dimensional 
quasi-linear hyperbolic systems. 
This collection of systems is enumerated by 
the genus g=0,1,2,\dots of the corresponding hyperelliptic
  Riemann surface. Each of these  systems 
 can be integrated by the so called  hodograph transform introduced by  Tsarev.
A key  step in the integration process is the solution
of the Tsarev linear overdetermined system. 
For each  $g>0$, we construct the unique solution of the Tsarev system,
which matches the genus $g+1$ and $g-1$ solutions on the 
transition  boundaries.   
Next  we  characterize initial data such that the solution of
the Whitham equations  has genus $g\leq N$, $N>0$,  for
all real $t\geq 0$ and $x$.
\end{abstract}
\vskip 10pt
\noindent

\section{Introduction}
The Whitham equations are a collection of quasi-linear hyperbolic systems 
of the form \cite{W},\cite{FFM},\cite{LL}
\begin{equation}
\dfrac{\partial u_i}{\partial t}-\lambda_i(u_1,u_2,\dots,u_{2g+1})\dfrac{
\partial u_i}{\partial x}=0,\quad x,t,u_i\in\Rn,\;\;\;i=1,...,2g+1,\;
\;\;g= 0,1,2,\dots,
\label{whithamg}
\end{equation}
with the ordering $u_1>u_2>\dots>u_{2g+1}$. 
For a given $g$ the system (\ref{whithamg}) is called $g$-phase
Whitham equations. Because of the diagonal form of systems
(\ref{whithamg}) the dependent variables $u_1>u_2>\dots>u_{2g+1}$
 are called Riemann
invariants.  For $g>0$ the speeds  $\lambda_i(u_1,u_2,\dots,u_{2g+1})$,
 $i=1,2,\dots,2g+1$,
depend  through $u_1,\dots,u_{2g+1}$ on complete hyperelliptic integrals
 of genus $g$. 
For this reason the $g$-phase system is also called genus $g$ system.
 The zero-phase Whitham equation  has the form
\begin{equation}
\label{burgereq}
\dfrac{\partial u}{\partial t}-6u\dfrac{\partial u}{\partial x}=0\,,
\end{equation}
where we use the notation $u_1=u$.
\noindent
In this paper we study the  initial value problem of the Whitham equations
 for monotone  increasing
smooth ($C^{\infty}$)  initial data $u(x,t=0)=u_0(x)$, where 
 the range of $u_0(x)$, $x\in \Rn$,  is the interval $(a,b)$ and 
$-\infty\leq a<b\leq +\infty$.

\noindent
 The initial value problem consists of the
following. We consider the evolution on the $x-u$ plane
 of the initial curve $u(x,t=0)=u_0(x)$ according to the zero-phase
equation (\ref{burgereq}). 
The solution $u(x,t)$ of (\ref{burgereq}) with the initial 
data $u_0(x)$  is given by the characteristic equation 
\begin{equation}
\label{zp}
x=-6tu+f(u)
\end{equation}
where $f(u)$ is the inverse function of $u_0(x)$.
The solution $u(x,t)$ in (\ref{zp})  is globally well 
defined only for $0\leq t<t_0$, where 
$t_0=\frac{1}{6}\min_{u\in\Rn}[f^{\prime}(u)]$ is the time of gradient 
catastrophe of  (\ref{zp}). 
 Near the point of gradient catastrophe and for  a short time  $t>t_0$, 
 the evolving curve is given  by
 a multivalued function with
three branches $b>u_1(x,t)>u_2(x,t)>u_3(x,t)>a$, which
 evolve according to  the one-phase Whitham equations. 
\begin{figure}[ht]
\centerline{\epsfig{figure=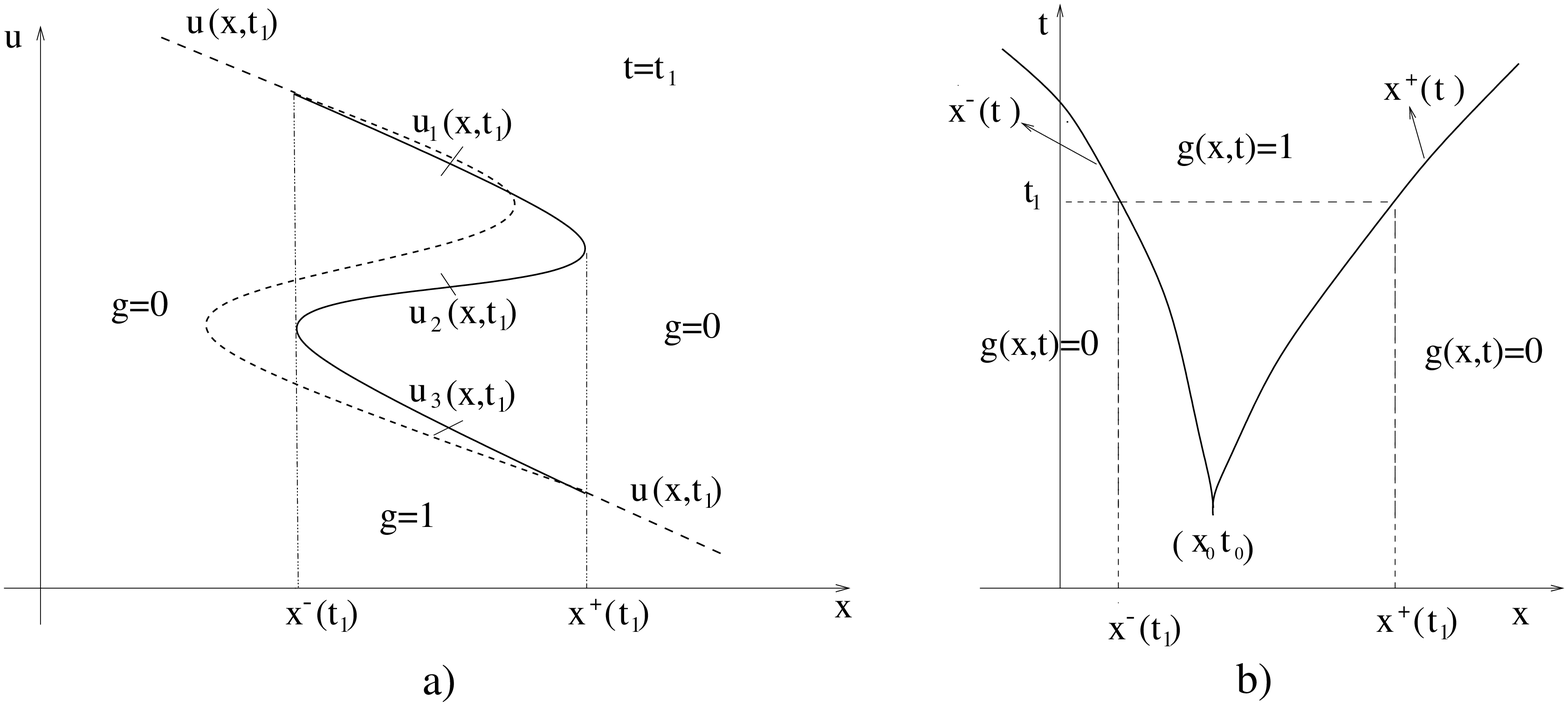,height=6cm}}
\caption{on   picture   a)  
the dashed line represents the formal solution
of the zero-phase equation the continuous line represents the
 solution of the one-phase equations. The solution
$(u_1(x,t)\,,\,u_2(x,t),u_3(x,t))$
of the one-phase equations  and the position of the boundaries 
$x^-(t)$, $x^+(t)$  are to be determined from the condition
$u(x^-(t),t)=u_1(x^-(t),t),\,\,$  
$u_x(x^-(t),t)=u_{1x}(x^-(t),t),\,\,$
$u(x^+(t),t)=u_3(x^+(t),t),\,$ $u_x(x^+(t),t)=u_{3x}(x^+(t),t)$,
where $u(x,t)$ is the solution of the zero-phase equation.
 Picture b) represents the functions $x^-(t)$ and $x^+(t)$  
on the $x-t$ plane.
\label{ivp} }
\end{figure}

Outside  the multivalued region the solution is given by  the zero-phase
solution $u(x,t)$ defined in (\ref{zp}). On the
phase transition boundary  the zero-phase solution and the one-phase 
solution are attached $C^1$-smoothly (see Fig\ref{ivp}).

\noindent
Since the Whitham equations are hyperbolic, other points 
of gradient catastrophe 
can appear in the  branches $u_1(x,t)>u_2(x,t)>u_3(x,t)$ themselves 
or in $u(x,t)$. 

\begin{figure}[tbh]
\centerline{\epsfig{figure=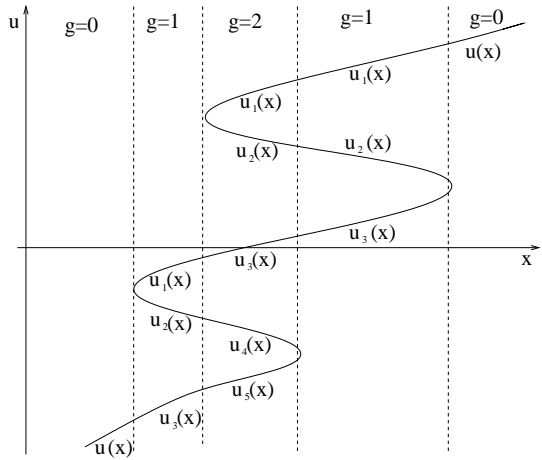,height=6cm}}
\caption{ the plot  on the $x-u$ plane   of the evolving multivalued  curve
for  fixed $t>0$.
\label{leeds4} }
\end{figure}

In general, for $t>t_0$,  the evolving curve is    given  by
 a multivalued function with and odd number of branches
 $b>u_1(x,t)>u_2(x,t)>\dots>u_{2g+1}(x,t)>a$, $g\geq 0$. 
These branches evolve according to the $g$-phase Whitham equations. 
The $g$-phase solutions  for
{\it different $g$}  must be glued together in order to produce a $C^1$-smooth
curve in the $(x,u)$ plane evolving smoothly with $t$ (see Fig.~\ref{leeds4}). 
The initial value problem of the Whitham equations is to determine, 
for almost all $t>0$ and $x$,  the phase
$g(x,t)\geq 0$ and the corresponding branches
 $u_1(x,t)>u_2(x,t)>\dots>u_{2g+1}(x,t)$
 from the initial data $x=f(u)|_{t=0}$. 
For example for the initial data $x=x_0+6t_0\,u+(u-u_0)^3$,  $t_0\geq 0$,  
the solution of the Whitham equations 
is of genus at most equal to one \cite{AN}\cite{FRT1}. It is of genus one
inside the cusp $-12\sqrt 3 (t-t_0)^{\frac{3}{2}}<x-x_0<
\frac{4}{3}\sqrt{\frac{5}{3}} (t-t_0)^{\frac{3}{2}}$ \cite{P}; 
it is of genus zero outside this cusp.
The point $x=x_0, t=t_0$ is the point of gradient catastrophe of the  
zero-phase solution.
The curve $x^-(t)=x_0-12\sqrt 3 (t-t_0)^{\frac{3}{2}}$, $t>t_0$, 
describes the locus of points where $u_1(x,t)=u_2(x,t)$. The curve
$x^+(t)=x_0+\frac{4}{3}\sqrt{\frac{5}{3}} (t-t_0)^{\frac{3}{2}}$, $t>t_0$, 
 describes the locus of points where
$u_2(x,t)=u_3(x,t)$. 
For generic initial data it is not known whether
the genus of the solution of the Whitham equations is bounded.
We say that the Cauchy   problem for the Whitham equations has  a global
solution if the genus $g<\infty$ for all $x$ and $t\geq 0$.

\noindent
Using the geometric-Hamiltonian structure \cite{DN} of the Whitham
equations,   Tsarev \cite{T} showed that these equations 
can be locally integrated by
a generalization of the method of  characteristic.  
Namely he proved that if
the functions  $w_i=w_i(u_1,u_2,\dots,u_{2g+1})$, $i=1,\dots,2g+1$,  
solve the linear over-determined system
\begin{equation}
\label{tsarev0}
\dfrac{\partial w_i}{\partial u_j}=\dfrac{1}{\lb_i-\lb_j}\dfrac{\partial \lb_i}{\partial u_j}[w_i-w_j],\quad
i,j=1,2,\dots,2g+1,\;\;i\neq j,
\end{equation}
where $\lb_i=\lb_i(u_1,u_2,\dots,u_{2g+1}),\; i=1,\dots, 2g+1$, are 
the speeds in  (\ref{whithamg}), then 
 the solution $\vu(x,t)=(u_1(x,t),u_2(x,t),\dots,u_{2g+1}(x,t))$ 
of the so called  hodograph transform
\begin{equation}
\label{ch0}
x=-\lambda_i(\vu)\,t+w_i(\vu)
\,\quad i=1,\dots ,2g+1\,,
\end{equation}
satisfies system (\ref{whithamg}). Conversely, any solution
$(u_1(x,t),u_2(x,t),\dots, u_{2g+1}(x,t))$  of (\ref{whithamg}) can be obtained in this
way.

\noindent
Tsarev theorem  relies on two factors:

a) the existence of a solution of the linear over-determined system
(\ref{tsarev0});

b) the existence of a real solutions 
$u_1(x,t)>u_2(x,t)>\dots>u_{2g+1}(x,t)$  of the hodograph 
transform (\ref{ch0}).

In this paper we solve completely problem a), namely we build for any 
smooth monotone increasing initial data $x=f(u)|_{t=0}$ the solution of system 
(\ref{tsarev0}) for any $g\geq 0$. The solution $w_i(\vu)$, $i=1,\dots,2g+1$, 
of genus $g$ satisfies some natural boundary conditions which guarantee 
its uniqueness. 

\noindent
Regarding problem b) we characterize initial data such that the solution of the
Whitham equations exists only for  $g\leq N$ where $N$ is a positive  integer.
Namely we show that if the initial data satisfies the condition
\begin{equation}
\label{derivative}
\dfrac{d^{2N+1}}{du^{2N+1}}f(u):= f^{(2N+1)}(u)>0, 
\quad 1\leq N\in\Nn
\end{equation}
for  all real $u$  belonging to the domain of $f$ except one point, 
then the  solution of the
Whitham equations has genus at most $N$ for any $x$ and $t\geq 0$.
For $N=1$ this result has already been proved by Tian 
\cite{FRT1}.

\noindent
The investigation of the initial value problem of the Whitham equations 
was initiated by Gurevich and  Pitaevskii \cite{GP}. 
In the case $g\leq 1$ they solved the initial value
problem of system (\ref{whithamg}) for step-like initial data and studied 
numerically the case of cubic initial data.

The initial value problem of the Whitham equations was deeply studied by Tian.
In \cite{FRT1} he  constructed the general solution of the Tsarev 
system (\ref{tsarev0}) for $0\leq g\leq 1$ and
for smooth monotone increasing
 initial data. He also proved the solvability of 
the hodograph  transform (\ref{ch0})
for $0\leq g\leq 1$. In \cite{FRT2} he obtained the solution of the Tsarev 
system for $g> 0$ for polynomial initial data.  

\noindent
The equations (\ref{whithamg})  were found by Whitham \cite{W} in the
  single phase case $g=1$  
and more generally by Flaschka, Forest and McLaughlin \cite{FFM} in the
multi-phase case. The Whitham equations were also found  in \cite{LL}  when 
 studying  the zero dispersion limit of the Korteweg de Vries equation.
The hyperbolic nature of the equations was found by Levermore \cite{L}.

\noindent
This paper is organized as follows.

\noindent
In Sec. 2 we give some background about Abelian differentials on 
hyperelliptic Riemann surfaces.

\noindent
In Sec. 3 we describe the Whitham equations. 

\noindent
In Sec. 4 we build the solution of the Tsarev
system (\ref{tsarev0}) for smooth monotone increasing initial data.

\noindent
In Sec. 5 we show that under the hypothesis (\ref{derivative})
 the hodograph transform is   solvable  for $g\leq N$. 

\noindent
In Sec. 6 we draw the conclusions.


\section{Riemann surfaces and Abelian differentials: notations and
 definitions}
\setcounter{equation}{0}
Let
\begin{equation}
\cs_g:=\left\{P=(r,\mu),\;\mu^2=\prod_{j=1}^{2g+1}(r-u_j)\right\}\,,
\quad u_1>u_2>\dots>u_{2g+1}\,,\;\;u_i\in\Rn\,,
\end{equation}
be the hyperelliptic Riemann surface of genus $g\geq 0$. We shall use the 
standard
representation of $\cs_g$ as a two-sheeted covering of $C\Pn^1$ with  cuts
along the intervals
\[
[u_{2k},u_{2k-1}],\quad k=1,\dots,g+1,\quad u_{2g+2}=-\infty\,.
\]
\noindent
We choose the  basis $\{\alpha_j,\beta_j\}_{j=1}^g$ of 
the homology group $H_1(\Gamma_g)$ so that $\alpha_j$ lies fully on the
  upper sheet
and encircles clockwise the interval $[u_{2j},u_{2j-1}]$, $j=1,\dots,g$,
 while $\beta_j$ emerges on the upper sheet on the cut  $[u_{2j},u_{2j-1}]$,
  passes anti-clockwise to the lower sheet trough the cut $(-\infty, u_{2g+1}]$
 and return to the initial point  through the lower sheet.

\noindent
The one-forms that are analytic on the closed Riemann surface $\cs_g$
except  for a finite number of points  are called Abelian differentials.

\noindent
We define on $\cs_g$ the following differentials \cite{S}:

\noindent
1) The canonical basis of holomorphic  one-forms or Abelian differentials
 of the first kind $\phi_1,\phi_2\dots\phi_g$:
\begin{equation}
\label{holo}
\phi_k(r)=\dfrac{r^{g-1}\gamma^k_1+r^{g-2}\ga^k_2+
\dots+\ga^k_g}{\mu(r)}dr\,,\quad k=1,\dots,g\,.
\end{equation}
The constants $\ga^k_i$ are uniquely determined by the 
 normalization  conditions  
\noindent
\begin{equation}
\int_{\alpha_j}\phi_k=\delta_{jk}\,,\quad j,k=1,\dots, g.
\end{equation}
We remark that an holomorphic differential having all its
$\alpha$-periods equal to zero is identically zero \cite{S}.

\noindent
2) The set $\sigma^g_k$, $k\geq 0$, $g\geq 0$, of Abelian differentials of the
second kind with a pole of order $2k+2$ at infinity, with asymptotic
behavior 
\begin{equation}
\label{sigma}
\sigma^g_k(r)=\left[r^{k-\frac{1}{2}}+O(r^{-\frac{3}{2}})\right]dr \quad 
\mbox{for
large}\;\; r
\end{equation}
  and normalized by the condition
\begin{equation}
\label{norm2}
\int_{\alpha_j}\sg^g_k=0,\quad j=1,\dots, g\,.
\end{equation}
We use the notation 
\begin{equation}
\label{dp}
\sg^g_0(r)=dp^g(r)\,,\quad\ 12\sg^g_1(r)=dq^g(r)\,\quad g\geq 0.
\end{equation} 
In literature the differentials $dp^g(r)$ and $dq^g(r)$ are called
quasi-momentum and quasi-energy respectively \cite{DN}.
The explicit formula for the differentials  $\sg^g_k$, $k\geq 0$, is given 
 by the expression
\beqa
\label{b6}
\sg^g_k(r)=\dfrac{P^g_k( r)}{\mu(r)}dr\;,
\quad P^g_k( r)= r^{g+k}+a^k_{1}
 r^{g+k-1}+a^k_{2} r^{g+k-2}\dots+a^k_{g+k}\,,
\eeqa
where  the coefficients 
$a^k_i=a^k_i(\vu)\;$,  $\vu=(u_1,u_2,\dots,u_{2g+1})$, $i=1,\dots,g+k,$ are uniquely determined by 
(\ref{sigma}) and (\ref{norm2}).

\noindent
3) The  Abelian differential of the
third kind  $\om_{qq_0}(r)$  with first order poles at the points 
$Q=(q,\mu(q))$ and $Q_0=(q_0,\mu(q_0))$  with residues
$\pm 1$ respectively. Its periods are normalized by the relation
\begin{equation}
\label{norm3}
\int_{\alpha_j}\om_{qq_0}(r)=0,\quad j=1,\dots,g\,.
\end{equation}

\subsection{Riemann bilinear relations}
\noindent
Let $\om_1$ and $\om_2$ be two Abelian differentials on the Riemann
surface $\cs_g$. 
 If all the residues of $\om_1$ and $\om_2$ are equal to zero,  then
the integrals
$d^{-1}\om_1$ and $d^{-1}\om_2$ do not have logarithm singularities 
on $\cs_g$.  If the differential $\om_1$  has non zero residue, then
its integral has logarithm singularities. Let  $s$ be the path
connecting the singular points of $d^{-1}\om_1$. 
We have the following relation.

\begin{equation}
\label{firstR}
\sum_{j=1}^g\left[\int_{\alpha_j}\om_1\int_{\beta_j}\om_2-
\int_{\alpha_j}\om_2\int_{\beta_j}\om_1\right]
+\int_s\Delta(d^{-1}\om_1)\om_2=2\pi i \sum_{\cs_g-s} \rres[ 
(d^{-1}\om_1)\om_2]\,,
\end{equation}
where $\Delta(d^{-1}\om_1)$ is the difference of the values of
$d^{-1}\om_1$ on the two  sides of the cut $s$ and the quantity $\sum_{\cs_g-s} \rres[ 
(d^{-1}\om_1)\om_2]$ 
 is the sum of the residues of the differential $(d^{-1}\om_1)\om_2$ on 
the cut surface $\cs_g-s$.
 This formula is known as the  Riemann bilinear period relation \cite{R}.

\noindent
Assuming $\om_1=\om_{qq_0}$ and $\om_2=\phi_k$ in (\ref{firstR}) we obtain
\begin{equation}
\label{i31}
\int_{\beta_k}\om_{qq_0}=2\pi i \int_{q_0}^{q}\phi_k\,,
\quad k=1,\dots,g\,. 
\end{equation}
Assuming $\om_1=\doq$ and $\om_2=\dop$ in (\ref{firstR})
we obtain
\begin{equation}
\label{int3}
\int_{p_0}^p\doq=\int_{q_0}^q\dop\,.
\end{equation}
Differentiating with respect to $p$ and $q$ the above expression we obtain
the identity
\begin{equation}
\label{diff3}
d_q[\doq(p)]=d_p[\dop(q)]\,,
\end{equation}
where $d_q$ and $d_p$ denote differentiation with respect to $q$ and
$p$ respectively.
From the expression (\ref{int3}) it follows that $\doq(r)$ is a many-value 
analytic function of the variable $q$.
The many-value character of $\doq(r)$ as a function of $q$ can be 
described by the equations
\begin{equation}
\label{MVP}
\int_{\alpha_k}d_q[\doq(r)]=0\,,\quad \int_{\beta_k}d_q[\doq(r)]=
2\pi i  \phi_k(r),\quad k=1,\dots,g\,,
\end{equation}

In the following we mainly use the normalized differential
 $\dozr$ which has  simple 
poles at the points $Q^{\pm}(z)=(z,\pm\mu(z))$
with residue $\pm 1$ respectively.

The differential  $\dozr$  is explicitly  given by the expression 
\begin{equation}
\label{CK}
\dozr=\dfrac{dr}{\mu(r)}\dfrac{\mu(z)}{r-z}-\sum_{k=1}^g\phi_k(r)
\int_{\alpha_k}\dfrac{dt}{\mu(t)}\dfrac{\mu(z)}{t-z}\,,
\end{equation}
where $\phi_k(r)$, $k=1,\dots,g$, 
is the  normalized basis of holomorphic differentials.
  $\dozr$  as a function of $z$, is an Abelian integral having poles of first
order at the points $Q^{\pm}(r)=(r,\pm\mu(r))$.
The periods of this integral are obtained from the relations (\ref{MVP})
\begin{equation}
\label{periods}
\int_{\alpha_j}d_z[\dozr]=0,\quad \int_{\beta_j}d_z[\dozr]=
4\pi i \phi_k(r)\,,\quad j=1,\dots g\,.
\end{equation}

We  apply
  the Riemann bilinear relation (\ref{firstR}) to  the differentials
  $\sg^g_m(r)$  and  $\dozr$ getting
\begin{equation}
\label{norm}
\begin{split}
\int_{Q^-(z)}^{Q^+(z)}\sg^g_m(\xi)=&-\res[r=\infty] 
[\dozr d^{-1}\sg^g_m(r)], 
\quad m=0,\dots,g, \\
=&-\dfrac{4}{2m+1}\left(-\mu(z)\e_{mg}+\sum_{k=1}^g\sum_{j=1}^{g} \gamma^k_{j}
\Gamma_{m+1-j} \int_{\alpha_k}\dfrac{dt}{\mu(t)}\dfrac{\mu(z)}{t-z}\right)\,.
\end{split}
\end{equation}
In the above formula $\e_{mg}=1$ for $m=g$ and zero otherwise, 
the coefficients  $\ga_j^k$ have been defined in 
(\ref{holo}) and the  $\Gamma_l$'s are the coefficients of 
the expansion for
  $\xi\ra \infty$ of
\begin{equation}
\label{GA}
\dfrac{1}{\mu(\xi)}=\xi^{-g-\frac{1}{2}}(\G_0+\dfrac{\G_1}{\xi}+
\dfrac{\G_2}{\xi^2}+\dots+\dfrac{\G_l}{\xi^l}+\dots).
\end{equation}
We define $\G_k=0$ for $k<0$. 

\noindent
Inverting (\ref{norm}) and introducing the quantities $N_j(z,\vu)=
\sum_{k=1}^g\gamma_j^k
\int_{\alpha_k}\dfrac{dt}{\mu(t)}\dfrac{\mu(z)}{t-z}  $ we obtain
\begin{equation}
\label{norm1}
\begin{pmatrix}
N_1(z,\vu)\\
N_2(z,\vu)\\
\dots\\
N_g(z,\vu)\\
-\mu(z)
\end{pmatrix}=
\begin{pmatrix}
\Gt_0&0&0&\dots&0\\
\Gt_1&\Gt_0&0&\dots&0\\
\dots&\dots&\dots&\dots&\dots\\
\Gt_{g-1}&\Gt_{g-2}&\dots&\Gt_0&0\\
\Gt_{g}&\Gt_{g-1}&\dots&\Gt_1&\Gt_0
\end{pmatrix}
\begin{pmatrix}
-\frac{1}{4}\int_{Q^-(z)}^{Q^+(z)}\sg^g_0(\xi)\\
-\frac{3}{4}\int_{Q^-(z)}^{Q^+(z)}\sg^g_1(\xi)\\
\dots\\
-\frac{2g-1}{4}\int_{Q^-(z)}^{Q^+(z)}\sg^g_{g-1}(\xi)\\
-\frac{2g+1}{4}\int_{Q^-(z)}^{Q^+(z)}\sg^g_{g}(\xi)
\end{pmatrix}
\end{equation}
where the $\Gt_k$'s are the coefficients of the expansion 
for $\xi\rightarrow\infty$ of
\begin{equation}
\label{Gt}
\mu(\xi)=\xi^{g+\frac{1}{2}}(\Gt_0+\dfrac{\Gt_1}{\xi}+
\dfrac{\Gt_2}{\xi^2}+\dots+\dfrac{\Gt_l}{\xi^l}+\dots).
\end{equation}
Using (\ref{norm1}) the differential $\dozr$ turns out to be given
by the relation
\begin{equation}
\label{dob}
\dozr=\dfrac{dr}{\mu(r)}\dfrac{\mu(z)}{r-z}-\sum_{k=1}^gN_k(z,\vu)\dfrac{r^{g-k}}{\mu(r)}dr\,.
\end{equation}
We remark that $\dozr$ is a multivalued function   of $z$ and it is regular at infinity.

\noindent 
From the relation (\ref{norm1}) we obtain the identity which will be useful later
\begin{equation}
\label{mu}
\mu(z)=\dfrac{1}{4}\sum_{k=1}^{g+1}(2k-1)\Gt_{g+1-k} \int_{Q^-(z)}^{Q^+(z)}\sg^g_{k-1}(\xi).
\end{equation}
The next proposition is also important for  our subsequent considerations.
\begin{prop}
The Abelian  differentials of the second kind  $\sg^g_k(r)$, $\,k\geq
0$, defined in (\ref{sigma})  
satisfy the relations 
\begin{equation}
\label{maini}
\sg^g_k(r)=\dfrac{1}{2}\resi\left[\dozr\,z^{k-\frac{1}{2}}dz\right]=-\dfrac{1}{2k+1}d_r\resi
\left[ \dorz\,z^{k+\frac{1}{2}}\right]\,,
\end{equation}
where $\dozr$ has been defined in (\ref{CK}),  
$\dorz$ is the normalized Abelian differential of the 
 third kind with   simple 
poles at the points $Q^{\pm}(r)=(r,\pm\mu(r))$
with residue $\pm 1$ respectively and $d_r$ denotes differentiation
with respect to $r$.
\end{prop}
\proof the differential $\resi\left[\dozr\,z^{k-\frac{1}{2}}dz\right]$ is
normalized because $\dozr$ is a normalized differential. Furthermore
\[
\resi\left[ \dozr\,z^{k-\frac{1}{2}} dz\right] = r^{k-\frac{1}{2}}dr+
O(r^{-\frac{3}{2}})dr\quad \mbox {for} \;r\ra \infty.
\]
 Therefore $ \resi\left[\dozr\,z^{k-\frac{1}{2}}dz\right]$ coincides with
the normalized Abelian differential of the second kind $\sg^g_k(r)$.
For proving the second equality in (\ref{maini}) we consider the
integral in the $z$ variable
\[
0=\oint_{C_{\infty}}d_z(\dozr\,z^{k+\frac{1}{2}})=
\oint_{C_{\infty}}z^{k+\frac{1}{2}}(d_z\dozr)+
\oint_{C_{\infty}}(k-\frac{1}{2})(\dozr\,z^{k+\frac{1}{2}})\,,
\]
where $C_{\infty}$ is a close contour around  the point at infinity.
Substituting the identity $d_z\dozr=d_r\dorz$ in the right hand side of   the above 
relation we obtain the second relation in (\ref{maini}).
\hfill$\square$
\vskip .2cm


\section{Preliminaries on the theory of the Whitham equations}
\setcounter{equation}{0}
The speeds $\lambda_i(u_1,u_2,\dots,u_{2g+1})$ 
of the $g$-phase Whitham equations (\ref{whithamg})  are given
by the ratio \cite{W},\cite{FFM}:
\begin{equation}
\label{b9}
\lambda_i(\vu)=\left.\dfrac{dq^g(r)}{dp^g(r)}\right|_{r=u_i}\;,\quad i=1,2,\dots, 
2g+1\,,
\end{equation}
where 
$dp^g(r)$ and $dq^g(r)$ have been defined in (\ref{dp}).
In the case $g=0$ 
\begin{equation}
\label{zerodiff}
dp^0(r)=\dfrac{dr}{\sqrt{r-u}}\,,\quad dq^0(r)=\dfrac{12r-6u}{\sqrt{r-u}}dr\,,
\end{equation}
so that  one obtains
the zero-phase Whitham equation (\ref{burgereq}).

For monotone increasing  smooth initial  data $x=f(u)|_{t=0}$,
 the solution of the zero-phase equation 
(\ref{burgereq}) is obtained by the method of characteristic \cite{W} 
and is given by the expression
\begin{equation}
\label{zerop}
x=-6tu+f(u)\,.
\end{equation}
The zero-phase solution is globally well defined only for $0\leq t <t_0$ where 
$t_0=\frac{1}{6}\min_{u\in\Rn}[f^{\prime}(u)]$ is the time of gradient 
catastrophe of the solution (\ref{zerop}).
 The breaking is cause by an inflection point 
in the initial data. 
For $t\geq t_0$ we expect to have single, double and higher phase solutions. 
For higher genus the Whitham equations can be locally integrated using
a generalization of the characteristic equation (\ref{zerop}).
We have the following theorem of Tsarev \cite{T}
\begin{theo}
\label{tsarevt}
If $w_i(\vu)$, $\vu=(u_1,u_2,\dots,u_{2g+1})$, solves the linear
over-determined system
 \begin{equation}
\label{tsarev}
\dfrac{\partial w_i}{\partial u_j}=a_{ij}(\vu)[w_i-w_j],\quad
i,j=1,2,\dots,2g+1,\;\;i\neq j,
\end{equation}
where
\begin{equation}
a_{ij}=\dfrac{1}{\lb_i-\lb_j}\dfrac{\partial \lb_i}{\partial u_j}\quad
i,j=1,2,\dots,2g+1,\;\;i\neq j,
\end{equation}
then the solution $(u_1(x,t),u_2(x,t),\dots,u_{2g+1}(x,t))$ 
of the hodograph transformation
\begin{equation}
\label{ch}
x=-\lambda_i(\vu)\,t+w_i(\vu)
\,\quad i=1,\dots ,2g+1\,,
\end{equation}
satisfies system (\ref{whithamg}). Conversely, any solution
$(u_1,u_2,\dots, u_{2g+1})$  of (\ref{whithamg}) can be obtained in this
way.
\end{theo}

To guarantee that the $g-$phase solutions for different $g$ are
attached continuously,    the following natural  boundary conditions must be imposed on
  $w_i(u_1,u_2,\dots,u_{2g+1})$, $i=1,\dots,2g+1$.

\noindent
When $u_l=u_{l+1}$, $\;1\leq l\leq 2g$,
\begin{equation}
\label{b1}
w_l^g(u_1,\dots,u_l,u_l,\dots,u_{2g+1})=
w_{l+1}^g(u_1,\dots,u_l,u_l,\dots,u_{2g+1})
\end{equation}
and for $1\leq i\leq  2g+1,\;i\neq l,\,l+1$
\beqa
\label{b2}  
w_i^g(u_1,\dots,u_l,u_l,\dots,u_{2g+1})=
w_i^{g-1}(u_1,\dots,\hat{u}_l,\hat{u}_l,\dots,u_{2g+1}).
\quad 
\eeqa 
The sup-scripts $g$ and $g-1$  in the $w_i$'s 
specify the corresponding genus 
and the hat denotes the variable that have been dropped.
When $g=1$ and  $u_2=u_3$ we have that 
\beqa
\label{b3}
\begin{array}{lll}
&w_1(u_1,u_3,u_3)=f(u_1)\\
  &w_2(u_1,u_3,u_3)=w_3(u_1,u_3,u_3)\,,
\end{array}
\eeqa
where $f(u)$ is the initial data. Similar conditions hold
true when $u_1=u_2$, namely
\beqa
\label{b4}
\begin{array}{lll}
&w_3(u_1,u_1,u_3)=f(u_3)\\
&w_1(u_1,u_1,u_3)=w_2(u_1,u_1,u_3).
\end{array}
\eeqa
We remark that the $\lb_i(\vu)$'s satisfy the boundary conditions 
(\ref{b1}-\ref{b2}) and for $g=1$ we have \cite{FRT2}
\[
\lb_1(u_1,u_3,u_3)=-6u_1,\quad 
\lb_2(u_1,u_3,u_3)=\lb_3(u_1,u_3,u_3)\,,
\]
and
\[
\lb_3(u_1,u_1,u_3)=-6u_3,\quad 
\lb_1(u_1,u_1,u_3)=\lb_2(u_1,u_1,u_3).
\]
The solution of the boundary value problem (\ref{tsarev}), (\ref{b1}-\ref{b4}) has been
obtained in \cite{FRT2,K} for  monotone increasing analytic initial data
of the form
\begin{equation}
\label{b18}
x=f_a(u)=c_0+c_1 u+\dots+c_{k}u^{k}+ \dots \,
\end{equation}
where we assume that only  a finite number of $c_k$ is different from
 zero. 
For such initial data the $w_i(\vu)$'s which satisfy
(\ref{tsarev}) and
 the boundary conditions (\ref{b1}-\ref{b4}) are given by the
expression
\cite{FRT2} 
\begin{equation}
\label{wi1}
w_i(\vu)=\left.\dfrac{ds^g(r)}{dp^g(r)}\right|_{r=u_i}\,,\quad i=1,\dots,2g+1.
\end{equation}
The differential $ds^g$ in (\ref{wi1}) is given by 
\begin{equation}
\label{ds}
ds^g(r)=\sum_{k=0}^{\infty} \dfrac{2^k k!}{(2k-1)!!} c_k \sigma^g_k(r)\,,
\end{equation}
 and the differentials $\sigma^g_k(r)$, $\,k\geq 0$  have been defined in 
(\ref{b6}).

\noindent
The solution (\ref{ch}) of the $g$-phase Whitham equations can also be 
written in an equivalent  algebro-geometric
 form  \cite{FRT2, K} namely
\begin{equation}
\label{b20}
 (-xdp^g(r)-tdq^g(r)+ds^g(r))\mid_{r=u_i}=0\;,\quad i=1,2\dots,2g+1\,,
\end{equation}
where $dp^g$, $dq^g$ and $ds^g$  
have been defined in (\ref{dp}) and (\ref{ds})
respectively. 

\noindent
The solution $u_1>u_2>\dots >u_{2g+1}$  of the $g$-phase Whitham
equations (\ref{whithamg})
 is  implicitly defined
 as a function of $x$ and $t$ by the equations  (\ref{ch}) or (\ref{b20}).
The solution is uniquely defined 
only for those 
$x$ and $t$ such that the functions $u_i(x,t)$ are real and
$\partial_xu_i(x,t)$, $i=1,\dots,2g+1$,  are not vanishing. 

\noindent
One of the problems in the theory of the Whitham equations is to determine
when  (\ref{ch}) or (\ref{b20})  are
 solvable for real $u_1,\dots,u_{2g+1}$ as functions
of $x$ and $t$.
This problem has been solved by Tian for $g\leq 1$.
\begin{theo}\cite{FRT1}
\label{Tian}
Consider a monotone increasing initial data $x=f(u)|_{t=0}$. Suppose that $u^*$
is the  inflection point of $f(u)$ that causes  the  breaking  of the 
zero-phase solution  (\ref{burgereq})  at $x=x_0, \;t=t_0$. Let be
$f^{\prime\prime\prime}(u)>0$ in a deleted neighborhood of $u=u^*$.
Then the one-phase Whitham equations has a solution $u_1>u_2>u_3$
within a cusp in the $x-t$ plane for a short time after the breaking
time  of the zero-phase solution. 
Furthermore this solution satisfies the boundary
conditions (\ref{b3}) and (\ref{b4}) on the cusp.
If the initial data satisfies the condition
$f^{\prime\prime\prime}(u)>0$ for all $u$ except  $u=u^*$, then 
the solution of the Whitham equations exists for all $t>0$.
The solution is of genus  one inside the cusp
$x^-(t)<x<x^+(t)$, $t>t_0$, where $x^-(t)<x^+(t)$ are two real
functions satisfying the condition $x^-(t_0)=x^+(t_0)=x_0$.
The solution is of genus zero outside the cusp $x^-(t)<x<x^+(t)$, $t>t_0$. 
\end{theo}


\section{Solution of Tsarev  system\label{slos}}
\setcounter{equation}{0}
In this section  we build  the solution of the boundary value problem
 (\ref{tsarev}),
(\ref{b1}-\ref{b4}) 
 for monotone smooth initial data and  we show that
the solution  obtained  is unique. We consider initial data of the form
$x=f(u)|_{t=0}$ where $f(u)$ is a monotone increasing function. The domain of
$f$ is the interval $(a,b)$ where  $-\infty\leq a<b\leq +\infty$,  and the range of
$f$ is  the real line $(-\infty,+\infty)$. 

\noindent
In order to obtain the solution of the boundary value problem
 (\ref{tsarev}),
(\ref{b1}-\ref{b4}) we need the following technical lemma.
\begin{lem}
The differential $ds^g(r)$ defined in (\ref{ds}) can be written in the form
\begin{equation}
\label{ds3}
ds^g(r)=2\mu(r)\left(\partial_r\Psi^g(r;\vu)+\sum_{k=1}^{2g+1}\partial_{u_k}
\Psi^g(r;\vu)\right)dr+\dfrac{R^{g}(r)}{\mu(r)}dr,
\end{equation}
where 
\begin{equation}
\label{psia}
\Psi^g(r;\vu)=-\resi\left[\dfrac{\cf(z)dz}{2\mu(z)(z-r)}\right], \quad
q_k(\vu)=-\resi\left[\dfrac{z^{g-k}\cf(z)dz}{2\mu(z)}\right],\;
k=1,\dots g,
\end{equation}
\begin{equation}
\label{abel}
\cf(z)=\int_{0}^{z}\dfrac{f_a(\xi)}{\sqrt{z-\xi}}d\xi,
\end{equation}
\begin{equation}
\label{R2ga}
R^{g}(r)=2\sum_{k=1}^{2g+1}\partial_{u_k}q_g(\vu)
\prod_{l=1,l\neq k}^{2g+1}(r-u_l)
+\sum_{k=1}^gq_k(\vu)
\sum_{l=1}^k(2l-1)\Gt_{k-l}P_{l-1}^g(r),
\end{equation}
and the polynomials $P_{l}^g(r)$, $l\geq 0$,  have been defined 
in (\ref{b6}), the  $\Gt_{k}$'s have been defined in (\ref{Gt}).
\end{lem} 
\proof using the second  identity in (\ref{maini}), we rewrite the differential
$ds^g(r)$ defined in (\ref{ds}) in the form
\begin{equation}
\label{ds2}
ds^g(r)=-d_r \left(\resi[\dorz \cf(z)]\right)\,,
\end{equation}
where $\dorz$  has been defined in (\ref{CK}) and $\cf(z)$
is the Abel transform defined in (\ref{abel}) of the analytic 
  initial data (\ref{b18}). The identity (\ref{ds2})
 can be checked straightforward. 
Using the explicit expression
of $\dorz$ in (\ref{dob}) we obtain
\begin{equation}
\label{ds31}
ds^g(r)=2d_r(\mu(r)\Psi^g(r;\vu))+\sum_{k=1}^gq_k(\vu)
\sum_{l=1}^k(2l-1)\Gt_{k-l}\sg_{l-1}^g(r),
\end{equation}
where $\Psi^g(r;\vu)$ and $q_k(\vu)$ have been defined in (\ref{psia}).

\noindent
From (\ref{psia}) we  get the relations
\begin{equation}
\label{psira}
\dfrac{\Psi^g(r;\vu)}{r-u_i}-\dfrac{\Psi^g(u_i;\vu)}{r-u_i}=
2\partial_{u_i}\Psi^g(r;\vu)
,\;\;\; 
2\partial_{u_i}q_g(\vu)=\Psi^g(u_i;\vu)
\end{equation}
and for $g=0$ we  define  $u_1=u$ and 
\[
2\partial_u q_0(u):=\Psi^0(u;u)=f_a(u).
\]
Using (\ref{psira})
we  transform the expression 
for $ds^g(r)$ in (\ref{ds31}) to the form (\ref{ds3}).
\hfill $\square$
\vskip .2cm

The relation  (\ref{ds3})   enables us to write the quantities 
$w_i(\vu)=\left.\frac{ds^g(r)}{dp^g(r)}\right|_{r=u_i}\,,\quad i=1,\dots,2g+1,$ in
(\ref{wi1}) in the form
\begin{equation}
\label{wia}
w_i(\vu)=
\dfrac{1}{P_0^g(u_i)}\left[2\partial_{u_i}q_g(\vu)
\prod_{l=1,l\neq i}^{2g+1}(u_i-u_l)+
\sum_{k=1}^gq_k(\vu)
\sum_{l=1}^k(2l-1)\Gt_{k-l} P_{l-1}^g(u_i)\right].
\end{equation}

\noindent
Observe that in the formula (\ref{wia})  all the information on the initial
data is contained in the functions  $q_k(\vu)$.
The functions $q_k=q_k(\vu)$, $k=1,\dots,g$,
solve the linear over-determined system \cite{FRT2} 
\beqa
\label{qka}
\begin{array}{rcl}
2(u_i-u_j)\dfrac{\partial^2 q_k(\vu)}{\partial u_i\partial u_j}&=&
\dfrac{\partial q_k(\vu)}{\partial u_i}-\dfrac{\partial
q_k(\vu)}{\partial u_j},
\quad i,j=1,\dots, 2g+1,\\
&&\\
q_k(\underbrace{u,u,\dots,u}_{2g+1})&=&\dfrac{2^{g-1}}{(2g-1)!!} u^{-k+\frac{1}{2}}\dfrac{d^{g-k}}{du}
\left(u^{g-\frac{1}{2}}f_a^{(k-1)}(u)\right),
\end{array}
\eeqa   
where 
 $f_a^{(k-1)}(u)$ is the $(k-1)th$ derivative of the polynomial initial data
$f_a(u)$. The function $\Psi^g(r;\vu)$ in (\ref{psia})  satisfies a similar linear 
 over-determined system.
\begin{theo}[First Main Theorem]
\label{theoT}
Let be $f(u)$ a  smooth monotone increasing function with domain $(a,b)$,
$-\infty\leq a<b\leq +\infty$. 
If $q_k=q_k(u_1,u_2,\dots,u_{2g+1})$, $1\leq k\leq g$, 
 is the symmetric
 solution of the linear over-determined system
\beqa
\label{qk}
\left\{\begin{array}{lll}
2(u_i-u_j)\dfrac{\partial^2 q_k(\vu)}{\partial u_i\partial u_j}=
\dfrac{\partial q_k(\vu)}{\partial u_i}-\dfrac{\partial
q_k(\vu)}{\partial u_j},
\quad i\neq j,\;i,j=1,\dots, 2g+1,\;\;g>0 &&\\
&&\\
q_k(\underbrace{u,u,\dots,u}_{2g+1})=F_k(u)&&\\
&&\\
F_k(u)=\dfrac{2^{(g-1)}}{(2g-1)!!} u^{-k+\frac{1}{2}}
\dfrac{d^{g-k}}{du^{g-k}}
\left(u^{g-\frac{1}{2}}f^{(k-1)}(u)\right),&&
\end{array}\right.
\eeqa   
with the ordering $u_1>u_2>\dots>u_{2g+1}$, then 
\begin{equation}
\label{wi}
w_i(\vu)=\dfrac{1}{P_0^g(u_i)}
\left[ 2\partial_{u_i}q_g(\vu)
\prod_{l=1,l\neq i}^{2g+1}(u_i-u_l)+
\sum_{k=1}^g q_k(\vu)
\sum_{l=1}^k(2l-1)\Gt_{k-l}P_{l-1}^g(u_i)\right], \quad i=1,\dots, 2g+1,
\end{equation}
solves  the boundary value problem (\ref{tsarev}),
(\ref{b1}-\ref{b4}). Conversely every solution of (\ref{tsarev}),
(\ref{b1}-\ref{b4}) can be obtained in this way.
\end{theo}
Before proving the theorem we show how to obtain the solution
 of  system (\ref{qk}) for generic
 smooth initial data. We follow the procedure in \cite{FRT2}. We start with the following lemma. 
\begin{lem}\cite{FRT2}
\label{Texist}
The system
\beqa
\label{Exist}
\left\{\begin{array}{cll}
2(z-y)p_{zy}&=&p_z-\rho p_y,\quad \rho>0\\
p(z,z)&=&s(z)
\end{array}\right.
\eeqa  
has, for any smooth initial data $s(z)$ one and only one
solution. Moreover,
the solution can be written explicitly
\begin{equation}
p(z,y)=\dfrac{1}{C_{\rho}}\int_{-1}^1\dfrac{  s(\frac{1+\mu}{2}z+\frac{1-\mu}{2}y)}
{\sqrt{1-\mu}}(1+\mu)^{\frac{\rho}{2}-1}d\mu
 \end{equation}
where
\begin{equation}
\label{Crho}
C_{\rho}=\int_{-1}^1 \dfrac{(1+\mu)^{\frac{\rho}{2}-1}}{\sqrt{1-\mu} }d\mu\,. 
\end{equation}
\end{lem}

Using the above lemma, the linear over-determined systems (\ref{qk})
can be integrated for any smooth initial data in the following way.
Suppose that $q_k(u_1,u_2,\dots,u_{2g+1})$ is a solution of
(\ref{qk}). 

Clearly 
$A_k(u_1,u_{2g+1})=
q_k(\underbrace{u_1,u_1,\dots,u_1}_{2g},u_{2g+1})$  satisfies
\beqa
\begin{array}{rcl}
2(u_1-u_{2g+1})\dfrac{\partial^2 A_k}{\partial u_1\partial u_{2g+1}}&=&
\dfrac{\partial A_k}{\partial u_1}-2g\dfrac{\partial
A_k}{\partial u_{2g+1}}\\
&&\\
A_k(u,u)&=&F_k(u)
\end{array}
\eeqa   
which by lemma~\ref{Texist} implies that
\begin{equation}
A_k(u_1,u_{2g+1})=\dfrac{1}{C_{2g}}\int_{-1}^1\dfrac{F_k(\frac{1+\xi_{2g}}{2}u_1+\frac{1-\xi_{2g}}{2}
u_{2g+1})}
{\sqrt{1-\xi_{2g}}}(1+\xi_{2g})^{g-1}d\xi_{2g}.
\end{equation}
For each fixed $u_{2g+1}$ the function 
$B_k(u_1,u_{2g},u_{2g+1})=q_k(\underbrace{u_1,\dots,u_1}_{2g-1},u_{2g},u_{2g+1})$ satisfies
\beqa
\label{qB}
\begin{array}{cll}
2(u_1-u_{2g})\dfrac{\partial^2 B_k}{\partial u_1\partial u_{2g}}&=&
\dfrac{\partial B_k}{\partial u_1}-(2g-1)\dfrac{\partial
B_k}{\partial u_{2g}}\\
&&\\
B_k(u,u,u_{2g+1})&=&A_k(u,u_{2g+1})
\end{array}
\eeqa   
Using again  lemma~\ref{Texist} we obtain
\begin{equation}
\begin{split} 
B_k(u_1,u_{2g},u_{2g+1})=&\dfrac{1}{C_{2g}C_{2g-1}}\int_{-1}^1\int_{-1}^1
d\xi_{2g}d\xi_{2g-1}
(1+\xi_{2g})^{g-1}(1+\xi_{2g-1})^{g-\frac{3}{2}}\times \\
&\\
&\dfrac{F_k(\frac{1+\xi_{2g}}{2}(\frac{1+\xi_{2g-1}}{2}u_1+
\frac{1+\xi_{2g-1}}{2}u_{2g})
\frac{1-\xi_{2g}}{2}u_{2g+1})}
{\sqrt{1-\xi_{2g}}\sqrt{1-\xi_{2g-1}}}
\end{split}
\end{equation}
Going on in the process of integration we obtain  the solution
$q_k(\vu)=q_k(u_1,u_2,\dots,u_{2g+1})$  of the boundary value problem
(\ref{qk}) namely
\begin{equation}
\label{fqk}
\begin{split} 
q_k(\vu)=&\dfrac{1}{C}\int_{-1}^1\int_{-1}^1\dots\int_{-1}^1
 d\xi_1d\xi_2\dots d\xi_{2g}(1+\xi_{2g})^{g-1}(1+\xi_{2g-1})^{g-\frac{3}{2}}
\dots(1+\xi_{3})^{\frac{1}{2}}(1+\xi_{1})^{-\frac{1}{2}}  \times\\
&\\
&\dfrac{F_k(\frac{1+\xi_{2g}}{2}(\dots(\frac{1+\xi_{2}}{2}(
\frac{1+\xi_{1}}{2}u_1+\frac{1-\xi_{1}}{2}u_2)+\frac{1-
\xi_{2}}{2}u_3)+\dots)+\frac{1-\xi_{2g}}{2}u_{2g+1})}
{\sqrt{(1-\xi_1)(1-\xi_2)\dots(1-\xi_{2g})}}
\end{split}
\end{equation}
where $C=\prod_{j=1}^{2g}C_{j}$ and $C_j$ has been defined in (\ref{Crho}).
When the initial data is of the form (\ref{b18}), the expression
(\ref{fqk}) for the $q_k(\vu)$'s 
is equivalent to (\ref{psia}). 
\begin{theo}
The boundary value problem (\ref{qk}) has one and only one
solution. The solution is symmetric and is given by (\ref{fqk}).
\end{theo}
\proof
Uniqueness follows from lemma~\ref{Texist} and the argument previous to
(\ref{fqk}). The boundary condition (\ref{qk}) is clearly satisfied.
The symmetry follows from the construction. Indeed in the process of
integration we can interchange the role of any of the variable $u_i$.
$\square$
\vskip .2cm
\noindent 
 We have the following relations.
\begin{lem}
The functions $F_k(u)$  and 
the  solutions
$q_k(\vu)$, $k=1,\dots,g$,  of the boundary value
 problem (\ref{qk}) 
 satisfy the
following relations.
\beqa
\label{relations}
\begin{array}{cll}
\partial_u F_k(u)&=&\dfrac{2g+1}{2}F_{k+1}(u)+u \partial_u F_{k+1}(u),\quad
k=1,\dots, g-1,\quad g>0\\
&&\\
\partial_{u_i}q_k(\vu)&=&\dfrac{1}{2}q_{k+1}(\vu)+u_i \partial_{u_i} 
q_{k+1}(\vu) \quad i=1.\dots,2g+1,\quad k=1,\dots,g-1\,\quad g>0.
\end{array}
\eeqa   
\end{lem}

\noindent 
{\bf Proof of Theorem~\ref{theoT} (First Main Theorem).}

\noindent
We consider the non trivial case where  $q_k(\vu)\not\equiv 0,\;
k=1,\dots g,$ and  $\partial_{u_j}q_g(\vu)\not\equiv 0$, $j=1,\dots,2g+1$. 

\noindent
 The proof consists of three parts.

\noindent
{\bf a)}{\it  The  $w_i(\vu)$'s defined in (\ref{wi}) satisfy (\ref{tsarev}).}

\noindent
Using the definition of $w_i(\vu)$ in (\ref{wi})  we 
have the following relation
\begin{equation}
\label{AA1}
\begin{split}
\partial_{u_j}w_i(\vu)=&
 2\dfrac{ \prod_{\overset{l=1}{l\neq i}}^{2g+1}(u_i-u_l) } { P_0^g(u_i) }\partial_{u_j}\partial_{u_i}q_g(\vu)-
 2\dfrac{ \prod_{\overset{l=1}{l\neq i}}^{2g+1}(u_i-u_l) } { P_0^g(u_i) }\partial_{u_i}q_g(\vu)
\left(\dfrac{ \partial_{u_j}P_0^g(u_i) } { P_0^g(u_i)}+\dfrac{1}{u_i-u_j} 
\right)\\
+&\partial_{u_j}\left( \sum_{l=1}^g(2l-1)\dfrac{P_{l-1}^g(u_i)}{P_0^g(u_i)} \sum_{k=l}^g q_k(\vu)\Gt_{k-l}
 \right), \quad i\neq j,\;\; i,j=1,\dots, 2g+1.
\end{split}
\end{equation}

The following identities hold
\cite{FRT2}
\begin{equation}
\label{x1}
\dfrac{\partial}{\partial_{u_j}}\dfrac{P_k^g(u_i)}{P_0^g(u_i)}=\dfrac{1}{\lb_i-\lb_j}\dfrac{\partial \lb_i}{\partial u_j}\left(\dfrac{P_k^g(u_i)}{P_0^g(u_i)}-\dfrac{P_k^g(u_j)}{P_0^g(u_j)}\right),\quad i\neq j, i,j=1,\dots,2g+1,\;\; k\geq 1
\end{equation}

\begin{equation}
\dfrac{1}{\lb_i-\lb_j}\dfrac{\partial \lb_i}{\partial u_j}=-\dfrac{\partial_{u_j}P_0^g(u_i)}{P_0^g(u_i)}-
\dfrac{1}{2}\dfrac{1}{u_i-u_j},\quad i\neq j,\; i,j=1,\dots,2g+1,\;\;
\end{equation}
where $\lb_i(\vu)$ has been defined in 
(\ref{b9}) and  $P_k^g(r)$ has been defined in (\ref{sigma}).

\noindent
Using the definition of $\Gt_k$ in (\ref{Gt}) it is easy to  verify that
\begin{equation}
\label{dGt}
\dfrac{\partial\Gt_k}{\partial u_j}=-\dfrac{1}{2}\sum_{m=1}^{k-1}\Gt_{k-m} u_j^{m-1}.
\end{equation}
Applying repeatedly   the relations (\ref{relations})   we obtain the following
expression for $\partial_{u_j}q_k(\vu)$:  
\begin{equation}
\label{x4}
\partial_{u_j}q_k(\vu)=\dfrac{1}{2}\sum_{m=1}^{g-k}q_{m+k}(\vu)u_j^{m-1}+u_j^{g-k}\partial_{u_j}q_g(\vu),\quad k=1,\dots,g-1.
\end{equation}

\noindent
From  (\ref{qk}) and (\ref{x1}-\ref{x4})  we can  write 
$\partial_{u_j}w_i(\vu)$ in (\ref{AA1}) in the form 
\begin{equation}
\nonumber
\begin{split}
\partial_{u_j}w_i(\vu)=&
\dfrac{\prod_{l=1,l\neq i}^{2g+1}(u_i-u_l)}{P_0^g(u_i)}
\dfrac{ \partial_{u_i}q_g(\vu)-\partial_{u_j}q_g(\vu) }{ u_i-u_j}\\
-& 2\dfrac{\prod_{l=1,l\neq i}^{2g+1}(u_i-u_l)}{P_0^g(u_i)}
\partial_{u_i}q_g(\vu)\left(
\dfrac{\partial_{u_j}P_0^g(u_i)}{P_0^g(u_i)}+\dfrac{1}{u_i-u_j}\right)\\
+&\sum_{l=1}^g(2l-1)\dfrac{P_{l-1}^g(u_i)}{P_0^g(u_i)}\left(
\dfrac{1}{2}\sum_{k=l}^{g-1}\Gt_{k-l}
\sum_{m=1}^{g-k}q_{m+k}(\vu)u_j^{m-1}+\sum_{k=l}^{g}\Gt_{k-l}u_j^{g-k}
\partial_{u_j}q_g(\vu)  \right)
\\
+&\sum_{l=1}^g(2l-1) \dfrac{P_{l-1}^g(u_i)}{P_0^g(u_i)}
\sum_{k=l}^g q_k(\vu) \left( -\dfrac{1}{2}\sum_{m=1}^{k-l}\Gt_{k-l-m} u_j^{m-1}  \right)
\\
+&\sum_{l=1}^g(2l-1)\,\dfrac{1}{\lb_i-\lb_j}\dfrac{\partial
\lb_i}{\partial u_j}\left(\dfrac{P_{l-1}^g(u_i)}{P_0^g(u_i)}-\dfrac{P_{l-1}^g(u_j)}{P_0^g(u_j)}\right) \sum_{k=l}^g q_k(\vu) \Gt_{k-l}  \quad i\neq j,\; i,j=1,\dots, 2g+1.
\end{split}
\end{equation}
Simplifying we obtain
\begin{equation}
\label{AA2}
\begin{split}
\partial_{u_j}w_i(\vu)=& -\dfrac{\prod_{l=1,l\neq i}^{2g+1}(u_i-u_l)}
{(u_i-u_j)P_0^g(u_i)}\partial_{u_j}q_g(\vu)
+2\dfrac{\prod_{l=1,l\neq i}^{2g+1}(u_i-u_l)}{P_0^g(u_i)}\partial_{u_i}q_g(\vu)\left(\dfrac{1}{\lb_i-\lb_j}\dfrac{\partial \lb_i}{\partial u_j} \right)\\
+&\sum_{l=1}^g(2l-1)\dfrac{P_{l-1}^g(u_i)}{P_0^g(u_i)}
\sum_{k=l}^{g}\Gt_{k-l}u_j^{g-k}\partial_{u_j}q_g(\vu)  \\
+&\dfrac{1}{\lb_i-\lb_j}\dfrac{\partial \lb_i}{\partial u_j} 
\sum_{l=1}^g(2l-1)\,\left(\dfrac{P_{l-1}^g(u_i)}{P_0^g(u_i)}-\dfrac{P_{l-1}^g(u_j)}{P_0^g(u_j)}\right) \sum_{k=l}^g q_k(\vu) \Gt_{k-l}  
\quad i\neq j,\; i,j=1,\dots, 2g+1.
\end{split}
\end{equation}
Adding and subtracting the quantity 
$\dfrac{1}{\lb_i-\lb_j}\dfrac{\partial \lb_i}{\partial u_j} w_j$ 
to (\ref{AA2}),
 we can reduce it to the form
\begin{equation}
\label{AA3}
\begin{split}
\partial_{u_j}w_i-&\dfrac{1}{\lb_i-\lb_j}\dfrac{\partial \lb_i}{\partial u_j}[w_i-w_j]=\partial_{u_j}q_g(\vu)\left(\dfrac{2}{\lb_i-\lb_j}\dfrac{\partial \lb_i}{\partial u_j}\dfrac{\prod_{l=1,l\neq j}^{2g+1}(u_j-u_l)}{P_0^g(u_j)}
 \right.\\
-&\left.\dfrac{\prod_{l=1,l\neq i}^{2g+1}(u_i-u_l)}{(u_i-u_j)P_0^g(u_i)} + \sum_{l=1}^g(2l-1)\dfrac{P_{l-1}^g(u_i)}{P_0^g(u_i)}
\sum_{k=l}^{g}\Gt_{k-l}u_j^{g-k}\right)
\end{split}
\end{equation}

The term in parenthesis
in the right hand side of (\ref{AA3}) does not depend on the initial
data
$f(u)$. It is identically zero for the  analytic initial data  (\ref{b18}) because
in such case the $w_i$'s satisfy  (\ref{tsarev}) \cite{FRT2}.
Therefore   we can conclude that
\begin{equation}
\partial_{u_j}w_i-\dfrac{1}{\lb_i-\lb_j}\dfrac{\partial \lb_i}{\partial u_j}
[w_i-w_j]=0
\end{equation}
for any smooth monotone increasing initial data $x=f(u)$.
\vskip 0.3cm
\noindent
{\bf b)}{\it  The  $w_i(\vu)$'s satisfy  the boundary conditions
  (\ref{b1}-\ref{b4}).} 

\noindent
 In the following we use the sup-script $g$ to
  denote the corresponding genus of the quantities we are referring
  to. We have the following relations. 
When $u_l=u_{l+1}=v$ for $l=1,\dots,2g$, the $\Gt_k$'s defined in (\ref{Gt})  satisfy
\begin{equation}
\label{z1}
\begin{split}
\Gt_k^g(u_1,\dots,u_{l-1},v,v,u_{l+2},\dots,u_{2g+1})=&\Gt_k^{g-1}(u_1,\dots,u_{l-1},u_{l+2},\dots,u_{2g+1})\\
-&v\Gt_{k-1}^{g-1}(u_1,\dots,u_{l-1},u_{l+2},\dots,u_{2g+1}),
\quad k\geq 1,\;g>1
\end{split}
\end{equation}
and the $q_k(\vu)$'s defined in (\ref{fqk}) satisfy
\begin{equation}
\label{z2}
\begin{split}
q_k^g(u_1,\dots,u_{l-1},v,v,u_{l+2},\dots,u_{2g+1})&-v
q_{k+1}^g(u_1,\dots,u_{l-1},v,v,u_{l+2},\dots,u_{2g+1})\\
&=q_k^{g-1}(u_1,\dots,u_{l-1},u_{l+2},\dots,u_{2g+1}),
\quad k=1,\dots,g-1,\;g>1.
\end{split}
\end{equation}
For $u_i\neq u_l=u_{l+1}=v$ we have that 
\begin{equation}
\begin{split}
\label{z3}
\partial_{u_i}q_{g-1}^{g-1}(u_1,\dots,u_{l-1},u_{l+2},\dots,u_{2g+1})-&
\dfrac{1}{2}q_g^g(u_1,\dots,u_{l-1},v,v,u_{l+2},\dots,u_{2g+1})=\\
&(u_i-v)\partial_{u_i}q_g^g(u_1,\dots,u_{l-1},v,v,u_{l+2},\dots,u_{2g+1}),
\end{split}
\end{equation}
which follows from (\ref{relations}).  When $u_l=u_{l+1}=v$
the polynomials $P^g_k(r)$'s defined in (\ref{b6})  satisfy the relation \cite{Fay}
\begin{equation}
\label{z4}
P_k^g(r)=(r-v)P_k^{g-1}(r),\quad k\geq 0\,.
\end{equation}
Using the relations (\ref{z1}-\ref{z4}) we have that for $i\neq l,\, l+1$, $\;i=1,\dots,2g+1$, 
\begin{equation}
\begin{split}
w_i^g(u_1,\dots,u_{l-1},v,v,u_{l+2},&\dots,u_{2g+1})= 2
\dfrac{\prod_{k=1,k\neq i,\,l,\, l+1}^{2g+1}(u_i-u_k)}{P_0^{g-1}(u_i)}
\partial_{u_i}q^{g-1}_{g-1}\\
+&\sum_{m=1}^{g-1}(2m-1)\dfrac{P^{g-1}_{m-1}(u_i)}{P_0^{g-1}(u_i)}
\sum_{k=m}^{g-1} q^{g-1}_k\Gt^{g-1}_{k-m}\\
+&\left( 
\sum_{m=1}^{g}(2m-1)\dfrac{P^{g-1}_{m-1}(u_i)}{P_0^{g-1}(u_i)}\Gt^{g-1}_{g-m}
-\dfrac{\prod_{k=1,k\neq i,\,l,\,l+1}^{2g+1}(u_i-u_k)}{P_0^{g-1}(u_i)}  \right)q_g^g, 
\end{split}
\end{equation}
which reduces to the form
\begin{equation}
\label{z5}
\begin{split}
w_i^g(u_1,\dots,u_{l-1},v,v,u_{l+2},&\dots,u_{2g+1})=
w_i^{g-1}(u_1,\dots,u_{l-1},u_{l+2},\dots,u_{2g+1})\\
+&\left( 
\sum_{m=1}^{g}(2m-1)\dfrac{P^{g-1}_{m-1}(u_i)}{P_0^{g-1}(u_i)}\Gt^{g-1}_{g-m}
-\dfrac{\prod_{k=1,k\neq i,\,l,\, l+1}^{2g+1}(u_i-u_k)}{P_0^{g-1}(u_i)}  \right)q_g^g.
\end{split}
\end{equation}
Using (\ref{mu}) the term in parenthesis in the right hand side of (\ref{z5})
turns out to be   identically 
zero. Therefore the boundary conditions (\ref{b2}) are satisfied for any smooth
monotone increasing initial data.

\noindent
Since the functions  $q_k(\vu)$'s in (\ref{fqk}) and the 
$\Gt_k(\vu)$'s defined in (\ref{Gt})
 are symmetric with respect to $u_1,u_2,\dots,
u_{2g+1}$ we immediately deduce from (\ref{z4}) that
\begin{equation}
\label{pp}
w_l^g(u_1,\dots,u_{l-1},v,v,u_{l+2},\dots,u_{2g+1})=
w_{l+1}^g(u_1,\dots,u_{l-1},v,v,u_{l+2},\dots,u_{2g+1})
\end{equation}
so that the boundary conditions (\ref{b1}) are satisfied (for a more
detailed analysis of this limit  see section~\ref{transition}).
When $g=1$ we deduce from (\ref{pp}) 
\[
w_1(u_1,u_1,u_3)=w_2(u_1,u_1,u_3)
\]
and  from (\ref{z3}-\ref{z4})
\[
w_3(u_1,u_1,u_3)=2(u_3-u_1)\partial_{u_3}q_1(u_1,u_1,u_3)+q_1(u_1,u_1,u_3).
\]
From (\ref{qk}) and (\ref{fqk})  we  get the relation 
\[
q_1(u_1,u_1,u_3)=f(u_3)+2(u_1-u_3)\partial_{u_3}q_1(u_1,u_1,u_3)
\]
so that 
\[
w_3(u_1,u_1,u_3)=f(u_3).
\]
An analogous result can be obtain when $u_2=u_3$,
 so that the boundary conditions 
(\ref{b3}-\ref{b4}) are satisfied.
\vskip 0.3cm
\noindent
{\bf c)} {\it  Uniqueness.}
\noindent
 We prove that when $f(u) \equiv 0$ then $w^g_i(\vu) \equiv 0$ for all
$b>u_1>u_2>\dots>u_{2g+1}>a$, for $1 \leq  i \leq  2g+1$ and for any
$g \geq 0$.
 The proof is by induction on $g$.
For $g=0$ the statement is satisfied.

For $g=1$ we repeat the arguments of \cite {FRT1}. We fix $u_2$ and we consider the equation (\ref{tsarev}) with
boundary condition (\ref{b3}-\ref{b4}), namely
\beqn
\begin{array}{rcl}
\dfrac{\partial w_1}{\partial u_3}&=&a_{13}[w_1-w_3]\\
&&\\
\dfrac{\partial w_3}{\partial u_1}&=&a_{31}[w_3-w_1]\\
w_1(u_1,u_2,u_2)&=&f(u_1)\equiv 0\\
w_3(u_2,u_2,u_3)&=&f(u_3)\equiv 0\,.
\end{array}
\eeqn
We can  regard the above equations as a first-order linear ordinary
differential equations with non homogeneous term. Integrating them 
we  obtain a couple integral equation. By standard contraction mapping
method it can be shown that when $f(u) \equiv 0$ 
this system has only zero solution,
i.e. $w_1=w_3\equiv 0$ for $(u_1,u_3)$ satisfying $b>u_1>u_2>u_3>a$. Because of
the arbitrariness of $u_2$,  $w_1$ and $w_3$ vanish as a function of
$(u_1,u_2,u_3)$ and therefore, by (\ref{tsarev}) so does $w_2(\vu)$.
Now we suppose the theorem true for genus $g-1$ and we proof  it 
 for genus $g$.
We fix $b>u_2>u_3>\dots>u_{2g}>a$ and we consider the equation
(\ref{tsarev})
for $w^g_1$    and $w^g_{2g+1}$ with
boundary condition (\ref{b1}-\ref{b2}), namely
\beqa
\begin{array}{rll}
\dfrac{\partial }{\partial u_{2g+1}}w^g_1&=&a_{1(2g+1)}[w^g_1-w^g_{2g+1}]\\
\dfrac{\partial}{\partial u_1} w^g_{2g+1}&=&a_{(2g+1)1}[w^g_{2g+1}-w^g_1]\\
w^g_1(u_1,u_2,\dots,u_{2g},u_{2g})&=&
w^{g-1}_1(u_1,u_2,\dots,u_{2g-1},\hat{u}_{2g},\hat{u}_{2g})\equiv 0\\
w^g_{2g+1}(u_2,u_2,\dots,u_{2g},u_{2g+1})&=&
w^{g-1}_{2g+1}(\hat{u}_2,\hat{u}_2,u_3,\dots,u_{2g},u_{2g+1})\equiv 0\,.
\end{array}
\eeqa
Repeating the arguments developed for genus $g=1$ we may conclude that
$w^g_1(\vu)=w^g_{2g+1}(\vu) \equiv0$,
for arbitrary $b>u_1>u_2>\dots>u_{2g+1}>a$. We then repeat the above
argument  fixing $b>u_1>u_3>\dots>u_{2g-1}>u_{2g+1}>a$ and
considering  the
equations (\ref{tsarev})  for $w^g_2(\vu)$ and $w^g_{2g}(\vu)$, namely
\beqa
\begin{array}{rll}
\dfrac{\partial }{\partial u_{2g}}w^g_2&=&a_{2(2g)}[w^g_2-w^g_{2g}]\\
\dfrac{\partial}{\partial u_2} w^g_{2g}&=&a_{(2g)2}[w^g_{2g}-w^g_2]\\
w^g_2(u_1,u_2,\dots,u_{2g-1},u_{2g+1},u_{2g+1})&=&
w^{g-1}_2(u_1,u_2,\dots,u_{2g-1},\hat{u}_{2g+1},\hat{u}_{2g+1})\equiv0\\
w^g_{2g}(u_1,u_1,u_3,\dots,u_{2g},u_{2g+1})&=&
w^{g-1}_{2g}(\hat{u}_1,\hat{u}_1,u_3,\dots,u_{2g},u_{2g+1})\equiv 0\,.
\end{array}
\eeqa
 It can be easily shown that also
$w^g_2(\vu)=w^g_{2g}(\vu)\equiv 0$ for  arbitrary 
$b>u_1>u_2>\dots>u_{2g+1}>a$. 
Repeating these arguments other $g-2$ times we conclude that
 $w^g_i(\vu) \equiv 0$
for $1\leq i\leq g,\;g+2\leq i \leq 2g+1$ and for arbitrary  
$b>u_1>u_2>\dots>u_{2g+1}>a$.
Applying  (\ref{tsarev}) and the boundary conditions (\ref{b1})-(\ref{b2})
we can prove that also $w^g_{g+1}(\vu)$ is
identically zero. The theorem is then proved.\hfill$\square$
\vskip .2cm
In the next section we consider the problem of reality of the solution
of the hodograph transform (\ref{ch}). 

\begin{rem}
The boundary conditions (\ref{b1}-\ref{b4}) guarantee the $C^1$-smoothness
of the solution of the Whitham equations. Indeed it can be proved that  
the $x$ derivatives   $\partial_x u_i(x,t)$, $i=1,\dots 2g+1$,  are continuous on the phase
transition boundaries.
\end{rem}


\section{An upper bound to  the genus of the solution}
\begin{figure}[tbh]
\centerline{\epsfig{figure=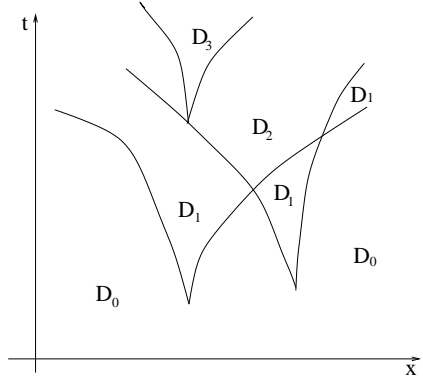,height=5cm}}
\caption{ An example of decomposition of the $x-t$ plane
\label{bbb} }
\end{figure}

\setcounter{equation}{0}
In this section we give an upper bound to the genus of the
solution of the Whitham equations for the  monotone increasing
smooth  initial data $x=f(u)|_{t=0}$. 
\begin{theo}[Second Main Theorem]
\label{main} 
If the monotone increasing smooth  initial data $f(u)$ satisfies the condition
\begin{equation}
\label{condition}
\dfrac{d^{2N+1}}{du^{2N+1}}f(u)>0, \quad 1\leq N \in\Nn,
\end{equation}
for all $u\in (a,b)$ except at one point,   then the  
solution of the
Whitham equations (\ref{whithamg})  has genus at most $N$.
\end{theo}
\proof the solution of the Whitham equations 
(\ref{whithamg}) for different $g\geq 0$  determines a
decomposition
of the $x-t$ plane, $t\geq 0$,   into a number of domains $D_g$ with
 g=0,1,2\dots, (see Fig.~\ref{bbb}).
 To the inner part of each domain $D_g$ it corresponds the
$g$-phase solution $b>u_1(x,t)>u_2(x,t)>\dots>u_{2g+1}(x,t)>a$
 of the Whitham equations (\ref{whithamg}). 
The common boundaries of the domains $D_g$, $g\geq 0$, 
 and $D_{g+n}$, $n\geq 1$, are 
the points of phase transition between
the $g$-phase solution and the $(g+n)$-phase solution.

We will show that each domain $D_g$, $g\leq N$, does not have a common 
boundary with any of  the domains $D_m$, $m>N$. Since the set of domains
$\{D_g\}_{g\leq N}$ is not empty because $D_0\neq \emptyset$ the
set $\{D_g\}_{g>N}$ must be empty. Indeed on the contrary
 the $x-t$ plane, $t\geq 0$, 
which is a connected set, would be  split  into  a number of domains 
whose union  forms a disconnected set.

\noindent
Before  determining  the  boundaries of the domains $D_g$, $g\geq 0$,
we need to study more in detail the hodograph transform
(\ref{ch}).

\begin{prop}
\label{zerophi}
Let us consider the polynomial
\begin{equation}
\label{zeta}
Z^g(r):=-xP^g_0(r)-12tP^g_1(r)+R^{g}(r),
\end{equation}
where    $R^0(r)=f(u)$ and  $R^{g}(r)$, $g>0$,  is given by the expression
\begin{equation}
\label{R2g}
R^{g}(r)=2\sum_{k=1}^{2g+1}\partial_{u_k}q_g(\vu)
\prod_{l=1,l\neq k}^{2g+1}(r-u_l)
+\sum_{k=1}^gq_k(\vu)
\sum_{l=1}^k(2l-1)\Gt_{k-l}P_{l-1}^g(r),
\end{equation}
with the polynomials $P^g_l(r)$, $l\geq 0$,  defined in (\ref{b6}) 
and the functions
$q_k(\vu)$, $k=1,\dots, g$,  defined in (\ref{qk}).
Then the hodograph transform (\ref{ch}) is equivalent, 
for $g>0$, to the equation
\begin{equation}
\label{zero}
Z^g(r)\equiv 0,\quad g>0.
\end{equation}
\end{prop}

\noindent
\proof
We observe that the $w_i(\vu)$'s  defined in (\ref{wi}) are given by the ratio 
$w_i(\vu)=\dfrac{R^{g}(u_i)}{P_0^g(u_i)}$, 
$i=1,\dots,2g+1$,  where $R^{g}(r)$ is the polynomial defined in  (\ref{R2g}).
Hence  we can write  the hodograph transform (\ref{ch}) in the
form
\begin{equation}
\label{SOL3}
[-xP^g_0(r)-12tP^g_1(r)+R^{g}(r)]_{r=u_i}=0, \quad i=1,\dots,2g+1. \;\;
\end{equation}
For $g>0$,  $\;Z^g(r)$ is a polynomial of degree $2g$  and
because of (\ref{SOL3})  
it must have at least $2g+1$ real zeros. Therefore it is identically
zero. Hence for $g>0,\;$  (\ref{zero}) is
equivalent to (\ref{SOL3}) and  (\ref{ch}).
\hfill $\square$
\vskip .2cm
In the following analysis we give some conditions 
for the existence of a real solution
$b>u_1(x,t)>u_2(x,t)>\dots, u_{2g+1}(x,t)>a$ of the 
hodograph transform (\ref{ch}). 
\noindent
For the purpose let us consider the function $\Psi^g(r;\vu)$ 
 which   solves the boundary value problem
\beqa
\label{bpsi}
\left\{\begin{array}{lll}
&&\dfrac{\partial}{\partial u_i}\Psi^g(r;\vu)-\dfrac{\partial}{\partial u_j}
\Psi^g(r;\vu)=2(u_i-u_j)\dfrac{\partial ^2}{\partial u_i\partial u_j}
\Psi^g(r;\vu),\quad i\neq j,\;\;i,j=1,\dots 2g+1\\
&&\dfrac{\partial}{\partial r}\Psi^g(r;\vu)-2\dfrac{\partial}{\partial u_j}
\Psi^g(r;\vu)=2(r-u_j)\dfrac{\partial ^2}{\partial r\partial u_j}
\Psi^g(r;\vu),\quad \;j=1,\dots 2g+1\\
&&\Psi^g(r;\underset{2g+1}{\underbrace{r,\dots,r}})=
\dfrac{2^{g}}{(2g+1)!!}f^{(g)}(r)
\end{array}\right.
\eeqa
where $f^{(g)}(r)$ is the $g$th derivative  of the
smooth monotone increasing initial data $f(u)$.
From the results of Sec.~\ref{slos} we are able to integrate (\ref{bpsi})
obtaining
\begin{equation}
\label{psi}
\begin{split} 
\Psi^g(r;\vu)=&\dfrac{1}{K}\int_{-1}^1\int_{-1}^1\dots\int_{-1}^1
 d\xi_1d\xi_2\dots d\xi_{2g+1}(1+\xi_{2g+1})^g(1+\xi_{2g})^{g-\frac{1}{2}}
\dots(1+\xi_2)^{\frac{1}{2}}\times\\
&\\
&\dfrac{ f^{(g)}(\frac{1+\xi_{2g+1}}{2}(\dots(\frac{1+\xi_{2}}{2}(
\frac{1+\xi_{1}}{2}r+\frac{1-\xi_{1}}{2}u_1)+\frac{1-
\xi_{2}}{2}u_2)+\dots)+\frac{1-\xi_{2g+1}}{2}u_{2g+1}) }
{ \sqrt{(1-\xi_1)(1-\xi_2)\dots(1-\xi_{2g+1})} }\\
\end{split}
\end{equation}
where $K=\dfrac{2^g}{(2g+1)!!}\prod_{j=2}^{2g+2}C_j$ and the $C_j$'s have been defined in (\ref{Crho}). The function $\Psi^g(r;\vu)$ is symmetric with respect to the variables
$b>u_1>u_2>\dots> u_{2g+1}>a$.
 For the initial data (\ref{b18}), the expression  of
$\Psi^g(r;\vu)$ in (\ref{psi}) is equivalent to (\ref{psia}).
\begin{prop}
\label{zphi}
Let us consider the function 
\begin{equation}
\label{Phi}
\Phi^g(r;\vu)=\partial_r\Psi^g(r;\vu)+\sum_{i=1}^{2g+1}
\partial_{u_i}\Psi^g(r;\vu),
\end{equation}
where $\Psi^g(r;\vu)$ has been defined in (\ref{psi}).
If  $u_1(x,t)>u_2(x,t)>\dots>u_{2g+1}(x,t)$ satisfy the Whitham equations 
(\ref{whithamg}) 
then the function $\Phi^g(r;\vu)$   has, in the $r$ variable, at
least one  real zero in each of the intervals
$(u_{2k},u_{2k-1}),\; k=1,\dots,g$, $g>0$.
\end{prop}
\proof
If the functions  $u_1(x,t)>u_2(x,t)>\dots>u_{2g+1}(x,t)$ 
satisfy the Whitham equations 
 then by proposition~\ref{zerophi} the polynomial 
$Z^g(r)\equiv 0$. Therefore
\begin{equation}
\begin{split}
0\equiv&\int_{\alpha_k}\dfrac{Z^g(r)}{\mu(r)}dr=\int_{\alpha_k}\dfrac{-xP^g_0(r)-12tP^g_1(r)+R^{g}(r)}{\mu(r)}dr\\
\label{ppp}
=&\int_{\alpha_k}2\dfrac{\sum_{i=1}^{2g+1}\partial_{u_i}q_g(\vu) 
\prod_{j=1,j\neq i}^{2g+1}(r-u_j)}{\mu(r)}dr,\quad k=1,\dots,g.\\
\end{split}
\end{equation}
In the third equality of (\ref{ppp})  we have used the fact that 
\[
\int_{\alpha_k}\dfrac{P^g_l(r)}{\mu(r)}dr=\int_{\alpha_k}\sg_l^g(r)=0,\quad l\geq 0,\;k=1,\dots,g,
\]
 because of the
normalization conditions (\ref{norm2}).
 The function $\Psi^g(r;\vu)$ satisfies the relations 
\begin{equation}
\label{psir}
\dfrac{\Psi^g(r;\vu)}{r-u_i}-\dfrac{\Psi^g(u_i;\vu)}{r-u_i}=
2\partial_{u_i}\Psi^g(r;\vu),\;\;\; 
2\partial_{u_i}q_g(\vu)=\Psi^g(u_i;\vu)
\end{equation}
which can be easily obtained from  (\ref{psi}).
Using (\ref{psir}) we can rewrite  the last term in (\ref{ppp}) in the form
\begin{equation}
\nonumber
\begin{split}
0=&\int_{\alpha_k}2\dfrac{\sum_{i=1}^{2g+1}\partial_{u_i}q_g(\vu) 
\prod_{j=1,j\neq i}^{2g+1}(r-u_j)}{\mu(r)}dr,\quad k=1,\dots,g,\\
=&2\int_{u_{2k}}^{u_{2k-1}}\mu(r)\left(\sum_{i=1}^g\dfrac{\Psi^g(r;\vu)}{r-u_i}-2\partial_{u_i}\Psi^g(r;\vu)\right)dr\\
=&-4\int_{u_{2k}}^{u_{2k-1}} \mu(r)\left(\partial_r \Psi^g(r;\vu)+
\sum_{i=1}^g\partial_{u_i}\Psi^g(r;\vu)\right)dr,\quad k=1,\dots,g,
\end{split}
\end{equation} 
where the last equality has been obtained integrating by parts.
 Using the definition of 
$\Phi^g(r;\vu)$  in (\ref{Phi}) we rewrite the above relation in the form
\begin{equation}
\label{normphi}
0=-4\int_{u_{2k}}^{u_{2k-1}} \mu(r)\Phi^g(r;\vu)dr,\quad k=1,\dots,g.
\end{equation}
Relation (\ref{normphi}) is satisfied only if the function $\Phi^g(r;\vu)$ changes sign
at least once in each of the intervals $(u_{2k},u_{2k-1})$, $k=1,\dots,g$.
\hfill$\square$
\vskip .2cm


In the following  we are going to determine the equations
which describe, on the $x-t\geq 0$ plane, 
 the boundary between the domains $D_g$ and $D_{g+1}$. 
  
The boundary between the domains $D_g$ and $D_{g+1}$
represents  a singular behavior in the solution of the $g$ or $(g+1)$-phase
equations. As we have  shown  in the example on  fig~\ref{ivp}, 
the boundary between the domains
$D_0$ and $D_1$ is described by  the curves $x^{\pm}(t)$ where
$u_2(x,t)=u_3(x,t)$ and $u_1(x,t)=u_2(x,t)$ respectively and by
 the point $x_0$, $t_0$ of gradient catastrophe
of the zero phase solution.
 
\noindent
In a similar way the generic boundary between the domains $D_g$ and $D_{g+1}$ is
described  by

\noindent
a) the curves $x_g^{\pm}(t)$   where  two Riemann invariants 
of the $(g+1)$-phase solution  coalesce;

\noindent
b) the  point of gradient catastrophe $x_c$, $t_c>0$  of the 
$g$-phase solution,   namely the point where one 
of the $2g+1$ Riemann invariants has a  vertical inflection point. 
The equations determining the point
of gradient catastrophe of the $g$-phase solution
 can also be obtained considering the limit of the
$g+1$ phase solution when three   Riemann invariants coalesce.

To treat  case a)  
  we  consider the Riemann surface $\cs_{g+1}$ of genus 
$g+1$ given by the equations
\[
\tilde{\mu}^2=(r-v-\sqrt\d )(r-v+\sqrt\d)\mu^2,\quad v\in\Rn\,,\;\;
\]
\[
\mu^2=
\prod_{j=1}^{2g+1}(r-u_j),\quad
b>u_1>u_2>\dots>u_{2g+1}>a,
\]
where $ v\neq u_j,\;\;j=1,\dots 2g+1$,  and $0<\d\ll 1$.
The Riemann invariants are the $2g+3$   variables
$\tilde{u}_1=v+\sqrt\d,\; \tilde{u}_2=v-\sqrt\d$, $\;u_1>u_2>\dots>u_{2g+1}$.
We suppose $\tilde{u}_1,\tilde{u}_2\neq u_j,\;j=1,\dots,2g+1$.
The hodograph transform (\ref{ch}) for these $2g+3$ variables
  has two different
behavior for  $\d\ra 0$ when $v$ belongs to one of the bands
\begin{equation}
\label{band}
(u_{2g+1},u_{2g})\cup(u_{2g-1},u_{2g-2})\cup\dots\cup(u_3,u_2)\cup(u_1,b)
\end{equation}
 or gaps
\begin{equation}
(a,u_{2g+1})\cup(u_{2g},u_{2g-1})\cup\dots\cup(u_4,u_3)\cup(u_2,u_1).
\label{gap}
\end{equation}
We call {\it leading edge}  of the phase transition boundary
the case in which $v$ belongs
to the bands.  We call {\it trailing edge}  of the phase transition boundary
the case in which $v$ belongs
to the gaps (\ref{gap}).

\begin{theo}
\label{LT}
The leading edge of the phase transition boundary between the 
$g$-phase solution and the $(g+1)$-phase solution is described  by 
the  system
\beqa
\label{y1}
\left\{\begin{array}{lll}
\Phi^g(v;\vu)-6t\,\e_{g0}=0&&\\
\partial_v\Phi^g(v;\vu)=0&&\\
x=\left[-12t\dfrac{P^{g}_1(r)}{P^{g}_0(r)}+
\dfrac{R^{g}(r)}{P^{g}_0(r)}\right]_{r=u_i}&&,\quad
i=1,\dots,2g+1,\; g\geq 0\\
\end{array}\right.
\eeqa
 where $\;v\in(u_{2j+1},u_{2j}),\;0\leq j\leq g,
\;u_{0}=+b$, the function  $\e_{g0}=1$ for $g=0$  and zero otherwise,  
 the function $\Phi^g(r;\vu)$ has been defined in (\ref{Phi}) 
and the polynomial
$R^g(r)$ has been defined in (\ref{R2g}).
We assume $(\partial_v)^2\Phi^g(v;\vu)\neq 0$ and $\Phi^g(u_i;\vu)\neq 0$, $i=1,\dots,2g+1$,  on the solution of 
(\ref{y1}).
\end{theo}

\begin{rem}
System (\ref{y1}) is a system of $2g+3$ equations in
$2g+4$ unknowns $x,t, v$ and $u_1>u_2>\dots>u_{2g+1}$. If
system (\ref{y1})  is uniquely solvable for real
$x, v$ and $u_1>u_2>\dots>u_{2g+1}$ as a function of $t\geq 0$, then a phase
transition between the $g$-phase solution and the $(g+1)$-phase solution
occurs. The curve $x_g^-=x_g^-(t)$ describes on the $x-t\geq 0$ plane 
the boundary between the domains $D_g$ and $D_{g+1}$ associated to the
leading edge. The conditions  $(\partial_v)^2\Phi^g(v;\vu)\neq 0$ and $\Phi^g(u_i;\vu)\neq 0$, $i=1,\dots,2g+1$,  on the solution of 
(\ref{y1}) exclude  higher order  degeneracy in the transition.  
Indeed in such case it can be proved that in a neighborhood of 
the solution $x(t)$, $v(t)$,
$u_1(t)>u_2(t)>\dots>u_{2g+1}(t)$ of (\ref{y1}) the   $(g+1)$-phase solution
is uniquely defined.
\end{rem}

\begin{theo}
\label{TT}
The trailing edge of the phase transition boundary between the 
$g$-phase solution and the 
$(g+1)$-phase solution is described  by the solution of the  system
\beqa
\label{T1}
\left\{\begin{array}{lll}
\Phi^g(v;\vu)-6t\,\e_{g0}=0&&\\
&&\\
\displaystyle\int_{v}^{u_{2j-1}}(\Phi^g(r;\vu)-6t\,\e_{g0})\mu(r)dr=0&&\\
&&\\
x=\left[-t\dfrac{P^{g}_1(r)}{P^{g}_0(r)}+
\dfrac{R^{g}(r)}{P^{g}_0(r)}\right]_{r=u_i},\quad
i=1,\dots,2g+1,\;g>0\\
\end{array}\right.
\eeqa
where $\;v\in(u_{2j},u_{2j-1}),\;1\leq j\leq g+1,
\;u_{2g+2}=a$ and the function $\Phi^g(r;\vu)$ has been defined in 
(\ref{Phi}). We assume $\partial_v\Phi^g(v;\vu)\neq 0$
and $\Phi^g(u_i;\vu)\neq 0$, $i=1,\dots,2g+1$,  on the solution
of (\ref{T1}). 
\end{theo} 
If
system (\ref{T1})  is uniquely solvable for real
$x, v$ and $u_1>u_2>\dots>u_{2g+1}$ as a function of $t\geq 0$, then a phase
transition between the $g$-phase solution and the $(g+1)$-phase solution
occurs. The curve $x_g^+=x_g^+(t)$ describes on the $x-t\geq 0$ plane 
the boundary between the domains $D_g$ and $D_{g+1}$ associated to the
trailing edge.

\noindent
The following theorem  enables one to 
determine a point of gradient catastrophe of the $g$-phase solution.
This point  can be  obtained  either as a limit of the $(g+1)$ phase solution 
when three Riemann invariants coalesce or imposing a vertical inflection
point on the $g$-phase solution.
\begin{theo}
\label{CC}
Let us consider the system
\beqa
\label{CCe}
\left\{\begin{array}{lll} 
\partial_r\Phi^g(r;\vu)|_{r=u_l}=0&&\\
\Phi^g(u_l;\vu)-6t\,\e_{g0}=0&&\\
x=\left[-12t\dfrac{P^{g}_1(r)}{P^{g}_0(r)}+
\dfrac{R^{g}(r)}{P^{g}_0(r)}\right]_{r=u_i},\quad
i=1,\dots,2g+1,\; g\geq 0\\
\end{array}\right.
\eeqa
where $\e_{g0}=1$ for $g=0$  and zero otherwise,    
the function $\Phi^g(r;\vu)$ has been defined in (\ref{Phi}).
When system (\ref{CCe})  is uniquely solvable for $x_c$, $t_c\geq 0$ and 
  $b>u_1(x_c,t_c)>u_2(x_c,t_c)>\dots> u_{2g+1}(x_c,t_c)>a$, 
then the $g$-phase solution has
a point of gradient catastrophe on the $u_l$ branch, $1\leq l\leq 2g+1$.
\end{theo}
To avoid higher order degeneracies in the transition
 we impose the condition
\begin{equation}
\label{third}
(\partial_r)^2\Phi^g(r;\vu(x_c,t_c))|_{r=u_l(x_c,t_c)}\neq 0. 
\end{equation}
Indeed we can prove that the above condition guarantees that the genus of 
solution of the Whitham equations increases at most by one in the
neighborhood of the point of gradient catastrophe.  Therefore it is
legitimate to consider the point of gradient catastrophe that solve (\ref{CCe})
and satisfies (\ref{third}) as a point of the boundary between the domains
$D_g$ and $D_{g+1}$. 
The condition (\ref{third}) in not essential in the genus $g=0$ case
as illustrated in Theorem~\ref{Tian}.

\noindent

\begin{rem}
We observe that both systems (\ref{y1}) and (\ref{T1}) in the 
limit $v\ra u_l$, $l=1,\dots,2g+1$, 
coincide with system (\ref{CCe}).
\end{rem}

\noindent
Theorems~\ref{LT},  \ref{TT} and \ref{CC}    characterize 
all types of boundaries between  the $g$-phase solution and the 
$(g+1)$-phase solution, namely the leading edge,
the trailing edge and the points of gradient catastrophe.
 We will prove these theorems  in the next section.

\vskip .3cm
{\bf Example 3.1}
For $x=u^k$, $k=3,5,7,\dots$,  the solution of the  Whitham equations
has genus at most equal to one \cite{AN},\cite{FRT1}. On the
$x-t$, $t\geq 0$, plane we have only the domains $D_0$ and $D_1$. 
The one phase solution is defined within the cusp 
 $x_-(k)\,t^{\frac{k}{k-1}}<x<x_+(k) \,
t^{\frac{k}{k-1}}$, where $x_-(k)<x_+(k)$ are two real constants and
$t>0$. The point $x=0, t=0$ is the point of gradient catastrophe of the
zero-phase solution.     
The constants $x_-(k)$ and $x_+(k)$ can be computed explicitly.
 From (\ref{psi}) and (\ref{Phi}) we obtain
\begin{equation}
\label{Phi0}
\Phi_0(r,u)=\dfrac{1}{2\sqrt{r-u}}\int_u^r\dfrac{f^{\prime}(\xi)d\xi}{\sqrt{r-\xi}}.
\end{equation}
On the leading edge where $u_1=u_2=v$ and $u_3=u$, $v>u$, 
 (\ref{y1})  is equivalent to the equations
\begin{equation}
\label{LL}
\int_u^v\dfrac{(\xi-u)f^{\prime\prime}(\xi)}{\sqrt{v-\xi}}\,d\xi=0,\quad
\int_u^v \dfrac{f^{\prime}(\xi)-6t}{\sqrt{v-\xi}}\,d\xi=0,\quad
x=-6tu+f(u).
\end{equation}
On the trailing edge where $u_1=u$, $u_2=u_3=v$, $u>v$, 
 system (\ref{T1}) is equivalent, for $g=0$  to 
\begin{equation}
\label{T00}
\int_v^u(f^{\prime}(\xi)-6t)\sqrt{v-\xi}\,d\xi=0,\quad
\int_u^v \dfrac{f^{\prime}(\xi)-6t}{\sqrt{v-\xi}}\,d\xi=0,\quad
x=-6tu+f(u).
\end{equation}
Equations (\ref{LL}) and (\ref{T00}) have already been obtained  
in \cite{FRT1}. 
Solving system (\ref{LL}) for $t$ when  $f(u)=u^k$, $k=3,5,7,\dots$
we obtain $x_-(k)$ \cite{P},\cite{G1}
\beqa
x_-(k)=-6\,\dfrac{k-1}{k}(2z_-(k)-1)
\left[\dfrac{6}{k}(1+2(k-1)z_-(k))\right]^{\frac{1}{k-1}}\,,
\quad k=3,5,7,\dots ,
\eeqa
where  $z_-(k)>1$ is the unique real solution of  
$F(-k+2,2,\frac{5}{2};z)=0$. Here  $F(a,b,c;z)$ is the
hypergeometric series.
The quantity $x_+(k)$ is obtained from (\ref{T00})  
\beqa
x_+(k)=2\,\dfrac{k-1}{k}(2z_+(k)-3)\left[\dfrac{2}{k}(3+2(k-1)z_+(k))\right]^
{\frac{1}{k-1}}\,,\quad k=3,5,7,\dots ,
\eeqa
where the  number $z_+(k)>1$ is the unique real solution of the equation
$F(-k+2,2,\frac{7}{2};z)=0$.
We give some numerical values: $x_-(3)=-12\sqrt3,\;x_-(5)=-16.85,\;
x_-(7)=-16.21,\;x_-(9)=-16.09,\;\\ x_+(3)=\frac{4}{3}\sqrt\frac{5}{3},\;x_+(5)=1.58,\;
x_+(7)=1.61,\;x_+(9)=1.72$.

\vskip 0.3cm

\noindent
Theorems~\ref{LT}, \ref{TT} and \ref{CC} 
can be generalized to the case of multiple phase
transitions. 
\begin{theo}
\label{Tmultiple}
The  transition boundary  between the domains $D_g$, $g\geq 0$  and $D_{g+n}$,
$n>1$,   having $n_1$ leading edges and $n_2$ trailing
edges and $n_3$ points of gradient catastrophes,  $n_1+n_2+n_3=n$, $n_1,n_2,n_3\geq 0$,
is described by the solution of the system
\beqa
\label{multiple}
\left\{\begin{array}{lll}
\partial_{v_k}\Phi^g(v_k;\vu)=0,\quad  k=1,\dots, n_1&&\\
\Phi^g(v_k;\vu)-6t\,\e_{g0}=0, \quad k=1,\dots, n_1&&\\
\Phi^g(y_l;\vu)-6t\,\e_{g0}=0,\quad l=1,\dots,n_2&&\\
\int_{y_l}^{u_{2j_l-1}}(\Phi^g(r;\vu)-6t\,\e_{g0})\mu(r)dr=0,\quad
l=1,\dots, n_2&&\\
\partial_r\Phi^g(r;\vu)|_{r=u_{j_m}}=0,\quad m=1,\dots,n_3&&\\
\Phi^g(u_{j_m};\vu)-6t\,\e_{g0}=0\quad m=1,\dots,n_3&&\\
&&\\
x=\left[-12t\dfrac{P^{g}_1(r)}{P^{g}_0(r)}+
\dfrac{R^{g}(r)}{P^{g}_0(r)}\right]_{r=u_i},\quad
i=1,\dots,2g+1,\; \\
\end{array}\right.
\eeqa
where $v_k\in (u_{2j_k},u_{2j_k+1})$, $0\leq j_k\leq g$, $k=1,\dots,n_1$
and $y_l\in (u_{2j_l-1},u_{2j_l})$,  $1\leq j_l\leq g+1$, $l=1,\dots,n_2$.
\end{theo}
\begin{rem}
The transition between the $g$-phase solution and the $(g+n)-$phase 
solution, $n>1$,  is highly non generic (cfr. Fig~\ref{multiple}). 
Indeed system  (\ref{multiple}) is a systems of $2n+2g+1$ equations in
$n_1+n_2+2g+3$ unknowns 
$v_1,\dots,v_{n_1}$, $\;y_1,\dots,y_{n_2}$, $u_1,\dots,u_{2g+1}$, 
$x$ and $t\geq 0$. Therefore if system (\ref{multiple}) admits a real solution,
$n+n_3-2$ variables  are functions of all  the others. 

\noindent
Theorem~\ref{Tmultiple} includes  all the degenerate cases 
that have been excluded in Theorems~\ref{LT}, \ref{TT} and \ref{CC}.
For example let us consider 
 a transition with a  double-leading edge described by  the  
variables $v_1 $, $v_2$ and $u_1>u_2>\dots>u_{2g+1}$ which solve (\ref{multiple}) with
$n_1=2$, $n_2=0,\; n_3=0$. 
In the limit $v_1\ra v_2=v$ we get   
 a single degenerate  leading edge. The variables $v$, $u_1>u_2>\dots>u_{2g+1}$
satisfy (\ref{LT}) and the equations $(\partial_v)^2\Phi^g(v;\vu)=0$ 
and $(\partial_v)^3\Phi^g(v;\vu)=0$.  Therefore we regard 
such degenerate single leading edge 
as a point of the boundary of  the domains $D_g$ and  $D_{g+2}$.  
\end{rem} 
\begin{figure}[tbh]
\centerline{\epsfig{figure=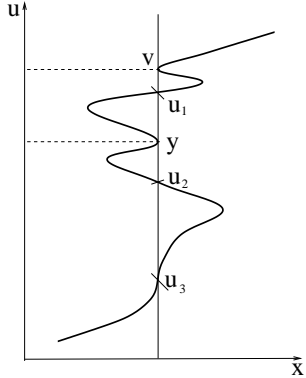,height=5cm}}
\caption{ an example of phase transition between  the one-phase   
solution and the $4$-phase  solution having one leading edge, one
trailing edge and one point of gradient catastrophe at $u_3$.
\label{multiplefig} }
\end{figure}

\begin{prop}
\label{propo}
If the domains $D_g$ and  $D_{g+1}$  have a common boundary,  then the 
  function 
\[
\Phi^g(r;\vu)-6t\e_{g0}
\]
 has  at least $g+2$ real zeros in the $r$ variable 
for $u_1(x,t)>u_2(x,t)>\dots>u_{2g+1}(x,t)$ satisfying (\ref{zero}) and
for  $t>0$.  
\end{prop}
\proof
 The domains  $D_g$ and
the $D_{g+1}$  have a common boundary if  one of the systems  (\ref{y1}),
 (\ref{T1}) or (\ref{CCe})  is solvable for some $t>0$.
We first consider the leading  edge.
For $g=0$ the statement is obvious from system (\ref{y1}). 
For $g>0$  system (\ref{y1}) can have 
 a solution if the function $\Phi^g(r;\vu)$
has a double zero at $r=v$ when $v$ belongs to the bands
(\ref{band}). 
Combining this observation with  proposition~\ref{zphi} we 
immediately obtain the statement.
As regarding the trailing edge, the theorem is obvious for $g=0$.
For $g>0$  and $v\in(a, u_{2g+1}) $  system (\ref{T1}) 
can be satisfied if the function $\Phi^g(r;\vu)$ has at least two zeros 
in the interval $(a, u_{2g+1}) $.  Combining this observation 
with  proposition~\ref{zphi} we  obtain the statement.
 When $v\in(u_{2j},u_{2j-1})$, $1\leq j\leq g$,
(\ref{normphi}) and (\ref{T1}) are satisfied if the function
$\Phi^g(r;\vu)$ has at least three real  zeros in the interval 
$(u_{2j},u_{2j-1})$, $1\leq j\leq g$, and  changes sign
at least once in each of the intervals $(u_{2k},u_{2k-1})$, 
$k=1,\dots,g$, and $k\neq j$.
 Therefore  $\Phi^g(r;\vu)$ has at least $g+2$ real zeros at 
the trailing edge. 
If the point of the boundary between the  domains  $D_g$ and
the $D_{g+1}$ corresponds to a point of gradient catastrophe of  the
$g$-phase solution the statement is obvious from system (\ref{CCe})
 and proposition~\ref{zphi}.
Therefore, on the phase transition boundary between the domains $D_g$ and
 $D_{g+1}$  the function $\Phi^g(r;\vu)$ has  at least $g+2$ real zeros in the
$r$ variable, $b>r>a$,  for $b>u_1(x,t)>u_2(x,t)>\dots>u_{2g+1}(x,t)>a$ and
for $t>0$.  . 
\hfill $\square$
\vskip .2cm
\noindent
Proposition~\ref{propo} can be generalized to multiple phase transitions.

\begin{prop}
\label{propom}
On the phase transition boundary between the domains  $D_g$  and
the $D_{g+n}$  the 
  function 
\[
\Phi^g(r;\vu)-6\,t\e_{g0}
\]
 has at least 
$g+2n$ real zeros in the $r$ variable  for 
$b>u_1(x,t)>u_2(x,t)>\dots>u_{2g+1}(x,t)>a$ and
for $t>0$.  
\end{prop}
The proof  is analogous to that of proposition~\ref{propo}.

\begin{lem}
\label{ngzero}
If the smooth initial data $x=f(u)|_{t=0}$ satisfies 
(\ref{condition}), then the  function  
\[
\Phi^g(r;\vu)
\]
has at most $2N-g$ real zeros (counting multiplicity) 
 in the $r$ variable for $0< g\leq 2N$ and for any real  $x$, $t\geq 0$
and $b>u_1>u_2>\dots>u_{2g+1}>a$.
\end{lem}
\proof
for proving the lemma we need the following elementary result.
If the real smooth function $\xi(r)$ satisfies the condition
\[
\dfrac{d^m}{dr^m}\xi(r)>0 ,\;0\leq  m\in\Nn, 
\] 
for all $r$ belonging to the domain of $\xi$,   
then $\xi(r)$ has at most $m$ real zero.

\noindent
Using   (\ref{condition}), (\ref{bpsi}) and (\ref{Phi})     we obtain
\begin{equation}
\label{phit}
\begin{split} 
&\dfrac{\partial^{2N-g}}{\partial r^{2N-g}}\Phi^g(r;\vu)=
\int_{-1}^1\dots\int_{-1}^1
 d\xi_1\dots d\xi_{2g+1}(1+\xi_{2g+1})^{2N}(1+\xi_{2g})^{2N-\frac{1}{2}}
\dots(1+\xi_2)^{2N-g+\frac{1}{2}}(1+\xi_1)^{2N-g}\\
&\\
&\times\dfrac{ f^{(2N+1)}(\frac{1+\xi_{2g+1}}{2}(\dots(\frac{1+\xi_{2}}{2}(
\frac{1+\xi_{1}}{2}r+\frac{1-\xi_{1}}{2}u_1)+\frac{1-
\xi_{2}}{2}u_2)+\dots)+\frac{1-\xi_{2g+1}}{2}u_{2g+1}) }
{2^{(2g+1)(2N-g)}\,K\, \sqrt{(1-\xi_2)(1-\xi_2)\dots(1-\xi_{2g+1})} }>0\\
\end{split}
\end{equation}
for $g>0$, for any real  $r$ and for any fixed real  $u_1>u_2>\dots >u_{2g+1}$ belonging to the
interval $(a,b)$.  
 Therefore  $\Phi^g(r;\vu)$  has at most $2N-g$ real zeros in the $r$ variable 
for  $0< g\leq 2N$ and for any fixed real  $b>u_1>u_2>\dots >u_{2g+1}>a$.   
\hfill $\square$
\vskip .2cm
\begin{lem}
\label{nzero}
If the smooth monotone increasing initial data $x=f(u)|_{t=0}$ satisfies 
(\ref{condition}), then the function 
\[
\Phi^0(r;u)-6t
\]
has at most $2N$ real  zeros (counting multiplicity)    in the $r$
variable, 
for any $t\geq 0$ and $u\neq u^*$, where $u^*$ is
  the unique solution of the equation  $f^{(2N+1)}(u^*)=0$.
For what  $\Phi^0(r;u^*)-6t$  is concerned we have two possibilities:
either it has at  most $2N$ real zeros (counting multiplicity)    
in the $r$ variable  for any $t\geq 0$ 
and  some of these  zeros are distinct,  or it
  has at most two real zeros in the $r$ variable  for any $t>0$
except for  $t=t^*\geq 0$  
 where  $r=u^*$ is a zero  of multiplicity higher  than $2N$. 
\end{lem}  
\proof
Since 
$f^{\prime}(u)\geq 0$ and because  of (\ref{condition}) we have 
two possibilities: either  $f^{\prime}(u)$  has
at most $N$ simple minima and some of these minima are distinct,
or  $f^{\prime}(u)$ is a non negative
  convex function with  a minimum at $u=u^*$, where  $u^*$ is  
the unique solution of the equation  
$f^{(2N+1)}(u^*)=0$. 
From the above considerations the lemma easily follows.
\hfill $\square$
\vskip .2cm
\noindent
We remark that when $f^{\prime}(u)$ is
a convex function,  we can apply 
Tian's Theorem~\ref{Tian}. 
Indeed in such case  we have  $f^{\prime\prime\prime}(u)>0$ for all $u\neq u^*$ 
and therefore the solution of the Whitham equations has genus at most one.
\vskip .2cm

\noindent 
We continue the proof of the second main theorem. In the following
 we exclude the case in which $f^{\prime}(u)$ is a   convex function. 
From proposition~\ref{propom}
we deduce that when 
    the domains $D_g$ 
and $D_{m}$, $m>g\geq 0$,   have a common boundary 
 the function $\Phi^g(r;\vu) -6t\e_{g0}$ defined in (\ref{Phi}) has at least  
$2m-g$ real zeros for some values of $x$ and $t$.
From lemma~\ref{ngzero}  the function 
$\Phi^g(r;\vu)-6t\e_{g0}$ defined in (\ref{Phi}) has at most $2N-g$ real zeros
for  $g\leq 2N$  when the
 initial data $f(u)$ satisfies 
(\ref{condition}). 
Therefore 
\begin{equation}
\label{c2}
2N-g\geq 2m-g\quad\mbox{or}\quad m\leq N.
\end{equation}
This shows that the
 set of domains $\{D_g\}_{0\leq g\leq N}$ does not have common
boundaries  with any of the domains in the set
  $\{D_m\}_{m>N}$. The second main theorem is proved. \hfill $\square$
\vskip .2cm

\subsection{Phase transitions\label{transition}\label{phaset}}
In this subsection we prove theorems~\ref{LT}, \ref{TT}, \ref{CC} and 
 (\ref{Tmultiple}).

\noindent
In the following we denote with a sup-script the genus of the 
quantity we are referring to. Whenever  we omit the sup-script we are referring
to genus $g$ quantity.

We denote  with  $\sg_k^{g+1}=\sg^{g+1}_{k}(r,\d,v)$
   the normalized Abelian differential
of the second kind with pole at infinity of order $2k+2$
  defined on the surface $\cs_{g+1}$
of genus $g+1$
\[
\tilde{\mu}^2=(r-v-\sqrt\d )(r-v+\sqrt\d)\mu^2,\quad v\in\Rn\,,\;\;
\]
\[
\mu^2=
\prod_{j=1}^{2g+1}(r-u_j),\quad
u_1>u_2>\dots>u_{2g+1},
\]
where $ v\neq u_j,\;\;j=1,\dots 2g+1$,  and $0<\d\ll 1$.
 We define the polynomials
\[
P_k^{g+1}(r,\d,v)=\tilde{\mu}(r)\dfrac{\sg_k^{g+1}(r,\d,v)}{dr}\,.
\]
The expression of the  polynomial
in (\ref{R2g}) for the Riemann invariants $\tilde{u}_1=v+\sqrt\d$,  $\tilde{u}_2=v-\sqrt\d$ and $u_1>u_2>\dots>u_{2g+1}$  reads
\begin{equation}
\begin{split}
R^{g+1}(r,\d,v)&=2\sum_{i=1}^{2g+1}\partial_{u_i}q^{g+1}_{g+1} 
\prod_{j=1,j\neq i}^{2g+1}(r-u_j)(r-\tilde{u}_1)(r-\tilde{u}_2)+
2\mu^2(r)(r-\tilde{u}_1)\partial_{\tilde{u}_2}q^{g+1}_{g+1}\\
+& 2\mu^2(r)(r-\tilde{u}_2)
\partial_{\tilde{u}_1}q^{g+1}_{g+1}+ 
\sum_{k=1}^{g+1}q^{g+1}_k
\sum_{l=1}^k(2l-1)
\tilde{\Gamma}^{g+1}_{k-l}P^{g+1}_{l-1}(r,\vu,\d)\,
\end{split}
\end{equation}
where  
$q^{g+1}_k=q^{g+1}_k(\tilde{u}_1,\tilde{u}_2,u_1,\dots,u_{2g+1})$, 
$\;k=1,\dots,g+1$ 
 and $\tilde{\Gamma}^{g+1}_{k}=\tilde{\Gamma}^{g+1}_{k}(\tilde{u}_1,\tilde{u}_2,u_1,\dots,u_{2g+1})$, $\,k\geq 0$.
We will  sometimes omit the explicit dependence of $\sg_k^{g+1}(r,\d,v)$,
 $ P_k^{g+1}(r,\d,v)$ and $R^{g+1}(r,v,\d)$ 
on  the parameters $v$ and $\d$. 

\noindent
{\bf Proof of Theorem~\ref{LT}}.

\noindent
We write the $(g+1)$-phase solution  (\ref{ch})
 for the variables
$\tilde{u}_1=v+\sqrt\d$, 
$\tilde{u}_2=v-\sqrt\d$, $\;u_1>u_2>\dots>u_{2g+1}$ in the form
\beqa
\label{wig1}
\left\{\begin{array}{lll}
0&=&\dfrac{1}{\tilde{u}_1-\tilde{u}_2}\left(\left[-12t\dfrac{P^{g+1}_1(\tilde{u}_1,\d)}{P^{g+1}_0(\tilde{u}_1,\d)}+
\dfrac{R^{g+1}(\tilde{u}_1,\d)}{P^{g+1}_0(\tilde{u}_1,\d)}\right]-
\left[-12t\dfrac{P^{g+1}_1(\tilde{u}_2,\d)}{P^{g+1}_0(\tilde{u}_2,\d)}+
\dfrac{R^{g+1}(\tilde{u}_2,\d)}{P^{g+1}_0(\tilde{u}_2,\d)}\right]\right),\\
&&\\
x&=&\left[-12t\dfrac{P^{g+1}_1(r,\d)}{P^{g+1}_0(r,\d)}+
\dfrac{R^{g+1}(r,\d)}{P^{g+1}_0(r,\d)}\right]_{r=\tilde{u}_2},\\
&&\\
x&=&\left[-12t\dfrac{P^{g+1}_1(r,\d)}{P^{g+1}_0(r,\d)}+
\dfrac{R^{g+1}(r,\d)}{P^{g+1}_0(r,\d)}\right]_{r=u_i},\quad
i=1,\dots,2g+1\\
\end{array}\right.
\eeqa
The reason to write the hodograph transform  (\ref{ch}) in the 
form (\ref{wig1}) is that
system (\ref{wig1}) is non degenerate at the phase transition 
$\tilde{u}_1=\tilde{u}_2$. 

\noindent
We  show that system (\ref{wig1}) reduces to system (\ref{y1}) when
$\tilde{u}_1=\tilde{u}_2$ or $\d=0$.

\noindent
It is clear from Theorem~\ref{theoT} and proposition~\ref{zerophi}
that  the last $2g+1$ equations in 
(\ref{wig1}) in the limit $\tilde{u}_1=\tilde{u}_2$ reduce to those of
system (\ref{y1}).

\noindent
For computing the limit $\tilde{u}_1\ra \tilde{u}_2$ of the first two
equations in (\ref{wig1}) 
 we first need to study the behavior of
 the differentials  $\sg_k^{g+1}(r,\d,v)$, $k\geq 0$, when 
  $\d\rightarrow 0$. When $v$ belongs to one of 
the bands (\ref{band}), 
 the expansion of
 $\sg_k^{g+1}(r,\d,v)$ for $\d\ra 0$  reads \cite{Fay}
\begin{equation}
\label{dplead}
\sg_k^{g+1}(r,\d,v)=\sg_k^g(r)+\dfrac{\d}{2}\sg_k^g(v)\partial_v\omega_v^g(r)
+O(\d^2)
\end{equation}
where  $\sg_k^g(r)$ is the normalized differential of the second 
kind defined on $\cs_g$
 with pole at infinity of order $2k+2$,  
$\;\sg_k^g(v)=\frac{\sg_k^g(r)}{dr}|_{r=v}$  and
$\partial_v=\dfrac{\partial}{\partial v}$.
The differential  $\omega_v^g(r)$ is  the normalized Abelian
differential of the third
kind with poles at the points $Q^{\pm}(v)=(v,\pm\mu(v))$ with residue
$\pm1$ respectively. The explicit expression of $\omega_v^g(r)$  has
been given  in (\ref{CK})
The differential $O(\d^2)/\d^2$ has a pole at $r=v$ of order $4$ and 
zero residue.

\noindent
From (\ref{dplead})  we can get the expansion of the polynomial 
$P_k^{g+1}(r,\d,v)=\tilde{\mu}(r)\dfrac{\sg_k^{g+1}(r,\d,v)}{dr}$, namely
\begin{equation}
\label{dpleadp}
P_k^{g+1}(r,\d,v)=(r-v) P_k^g(r)+\dfrac{\d}{2}\sg_k^g(v)\left(\mu^{\prime}(v)-(r-v)\sum_{k=1}^g r^{g-k}
N^{\prime}_k(v)\right)+O(\d^2),
\end{equation}
where  $P_k^g(r)$ has been defined in (\ref{b6}) and the $N_k(v)$'s have been defined in  (\ref{norm1}). 
From (\ref{dpleadp}) we can evaluate the following
\begin{equation}
\label{Plead}
\left.\dfrac{P_k^{g+1}(r,\d)}{P_0^{g+1}(r,\d)}\right|_{r=v\pm\sqrt\d}=
\dfrac{P_k^{g}(v)}{P_0^g(v)}\pm\sqrt\d\partial_v\left(\dfrac{P_k^{g}(v)}{P_0^g(v)}  \right)+O(\d).
\end{equation}

To evaluate the first two equations in (\ref{wig1}) at the point of
phase transition we need also the  following relations  
\begin{equation}
\label{qpsi}
\begin{split}
q^{g+1}_{g+1}(v,v,\vu)&=\Psi^g(v;\vu)\\
\left.\dfrac{\partial}{\partial \tilde{u}_i}q^{g+1}_{g+1}(\tilde{u}_1,\tilde{u}_2,\vu)
\right|_{\tilde{u}_1=\tilde{u}_2}&
=\dfrac{1}{2}\dfrac{\partial}{\partial v}\Psi^g(v;\vu),\quad i=1,2,\\
\left.\dfrac{\partial^2}{\partial \tilde{u}_1\partial \tilde{u}_2}q^{g+1}_{g+1}(\tilde{u}_1,\tilde{u}_2,\vu)\right|_{\tilde{u}_1=\tilde{u}_2}&
=\dfrac{1}{4}\dfrac{\partial^2}{\partial v^2}\Psi^g(v;\vu),
\end{split}
\end{equation}
where the function  $\Psi^g(r;\vu)$ has been defined in (\ref{psi}).
From  (\ref{z1}-\ref{z2}), (\ref{Plead}) and (\ref{qpsi})  we obtain 
\begin{equation}
\begin{split}
\left.\dfrac{R^{g+1}(\tilde{u}_i,\d)}{P_0^{g+1}(\tilde{u}_i,\d)}\right|_{\tilde{u}_1=\tilde{u}_2}&=\dfrac{2\mu^2(v)\partial_v\Psi^g(v;\vu)}{P_0^g(v)}+ 
\Psi^g(v;\vu)\sum_{l=1}^{g+1}(2l-1)
\tilde{\Gamma}^{g}_{g+1-l}\dfrac{P^{g}_{l-1}(v)}{P^g_0(v)} \\
&+ \sum_{k=1}^{g}q^g_k(\vu)
\sum_{l=1}^k(2l-1)
\tilde{\Gamma}^{g}_{k-l}\dfrac{P^g_{l-1}(v)}{P^g_0(v)},
\quad i=1,2.
\end{split}
\end{equation}
From  (\ref{mu}), (\ref{psir}) and using the definition of
$\Phi^g(r;\vu)$ in (\ref{Phi}) the above relation 
can be written in the  form 
\begin{equation}
\label{h1}
\left.\dfrac{R^{g+1}(\tilde{u}_i,\d)}{P_0^{g+1}(\tilde{u}_i,\d)}\right|_{\tilde{u}_1=\tilde{u}_2}=
\dfrac{2\mu^2(v)
\Phi^g(v;\vu)+R^{g}(v,\vu)}{P^{g}_0(v)}, \quad i=1,2,
\end{equation}
where $R^g(r,\vu)$ has been defined in (\ref{R2g}).
We need to consider also the quantity
\begin{equation}
\nonumber
\begin{split}
&\left[\dfrac{1}{\tilde{u}_1-\tilde{u}_2}\left(\dfrac{R^{g+1}(\tilde{u}_1,\d)}
{P^{g+1}_0(\tilde{u}_1,\d)}-
\dfrac{R^{g+1}(\tilde{u}_2,\d)}{P^{g+1}_0(\tilde{u}_2,\d)}\right)\right]_{\tilde{u}_1=
\tilde{u}_2}=
8\left.\left(\sqrt\d\dfrac{\mu^2(v+\sqrt\d)}{P_0^{g+1}(v+\sqrt\d,\d)}\partial_{\tilde{u}_1}\partial_{\tilde{u}_2}q^{g+1}_{g+1}(\tilde{u}_1,\tilde{u}_2,\vu)\right)\right|_{\d=0}\\
+&2\left.\left(\left(\dfrac{\mu^2(v+\sqrt\d)}{P_0^{g+1}(v+\sqrt\d,\d)}-\dfrac{\mu^2(v-\sqrt\d)}{P_0^{g+1}(v-\sqrt\d,\d)}\right)\partial_{\tilde{u}_2}q^{g+1}_{g+1}(\tilde{u}_1,\tilde{u}_2,\vu)\right)\right|_{\d=0}\\
+&\sum_{k=1}^{g}q^g_k(\vu)
\sum_{l=1}^k(2l-1)
\tilde{\Gamma}^{g}_{k-l}(\vu)\left.\left(\dfrac{1}{2\sqrt\d}\left(\dfrac{P^{g+1}_{l-1}(v+\sqrt\d,\d)}{ P^{g+1}_0(v+\sqrt\d,\d)}-\dfrac{P^{g+1}_{l-1}(v-\sqrt\d,\d)}{ P^{g+1}_0(v-\sqrt\d,\d)}\right)\right)
\right|_{\d=0}\\
+& q_{g+1}^{g+1}(v,v,\vu)\sum_{l=1}^{g+1}(2l-1)
\tilde{\Gamma}^{g}_{g+1-l}(\vu)\left.\left(\dfrac{1}{2\sqrt\d}\left(\dfrac{P^{g+1}_{l-1}(v+\sqrt\d,\d)}{P^{g+1}_0(v+\sqrt\d,\d)}-\dfrac{P^{g+1}_{l-1}(v-\sqrt\d,\d)}
{ P^{g+1}_0(v-\sqrt\d,\d)}\right)\right)
\right|_{\d=0}
\end{split}
\end{equation}
where we have used (\ref{z1}-\ref{z2}) to obtain the right hand side.
Using (\ref{mu}), 
(\ref{psir}), (\ref{Plead}) and (\ref{qpsi}) the above reduces to the form
\begin{equation}
\label{h2}
\left[\dfrac{1}{\tilde{u}_1-\tilde{u}_2}\left(\dfrac{R^{g+1}(\tilde{u}_1,\d)}
{P^{g+1}_0(\tilde{u}_1,\d)}-
\dfrac{R^{g+1}(\tilde{u}_2,\d)}{P^{g+1}_0(\tilde{u}_2,\d)}\right)\right]_{\tilde{u}_1=
\tilde{u}_2}=\dfrac{\partial}{\partial_v}\left(\dfrac{2\mu^2(v)
\Phi^g(v)+ R^{g}(v)}{P^{g}_0(v)}\right).
\end{equation}

From
(\ref{Plead}), (\ref{h1}) and (\ref{h2}), the system (\ref{wig1}) on the 
point of phase transition $\tilde{u}_1=\tilde{u}_2=v$
 reads
\beqa
\label{wg1}
\left\{\begin{array}{lll}
0&=&\dfrac{\partial}{\partial v}\left(\dfrac
{-12t P_1^g(v)+R^{g}(v,\vu)+2\mu^2(v)\Phi^g(v;\vu)}{P^g_0(v,\vu)}\right)\\
&&\\
x&=&\dfrac{-12t P_1^g(v,\vu)+R^{g}(v,\vu)+2\mu^2(v) \Phi^g(v;\vu)}{P^{g}_0(v,\vu)}\\
&&\\
x&=&\left[-12t\dfrac{P^{g}_1(r,\vu)}{P^{g}_0(r,\vu)}+
\dfrac{R^{g}(r)}{P^{g}_0(r,\vu)}\right]_{r=u_i},\quad
i=1,\dots,2g+1\\
\end{array}\right.
\eeqa

From proposition~\ref{zerop} the last $2g+1$ equations in (\ref{wg1})
are equivalent to the condition
\[
-xP^g_0(r)-12tP^g_1(r)+R^{g}(r)\equiv 0\,,\quad g>0.
\]
Therefore  system (\ref{wg1}) is equivalent to 
system (\ref{y1}) for $g>0$. For $g=0$ substituting (\ref{zerodiff}) in 
(\ref{wg1}) we obtain 
\beqa 
\label{y00}
\left\{\begin{array}{lll}
0= \partial_v(-6tu+f(u)+2(v-u)(\Phi_0(v,u)-6t))&&\\
&&\\
x=-6tu+f(u)+2(v-u)(\Phi_0(v,u)-6t)&&\\
&&\\
x=-6tu+f(u)&&\\
\end{array}\right.
\eeqa
 where  $u_1=u$ and  $f(u)$ is the initial data. It is clear that (\ref{y00}) 
is equivalent  to (\ref{y1}) for $g=0$.

\hfill $\square$
\vskip .2cm

\noindent
{\bf Proof of Theorem~\ref{TT}}.
 
For getting the equations determining the trailing edge we have to repeat all the
above calculation in a slightly different way.
We write the $(g+1)$-phase solution 
 for the variables
$\tilde{u}_1=v+\sqrt\d$, 
$\tilde{u}_2=v-\sqrt\d$, $\;u_1>u_2>\dots>u_{2g+1}$ in the form
\beqa
\label{ug1}
\left\{\begin{array}{lll}
0&=&\dfrac{-1}{\log(\frac{\tilde{u}_1-\tilde{u}_2}{2})^2(\tilde{u}_1-\tilde{u}_2)}\left(12t\left(\dfrac{P^{g+1}_1(\tilde{u}_2,\d)}{P^{g+1}_0(\tilde{u}_2,\d)}
-\dfrac{P^{g+1}_1(\tilde{u}_1,\d)}{P^{g+1}_0(\tilde{u}_1,\d)}\right)-
\dfrac{R^{g+1}(\tilde{u}_1,\d)}{P^{g+1}_0(\tilde{u}_1,\d)}+
\dfrac{R^{g+1}(\tilde{u}_2,\d)}{P^{g+1}_0(\tilde{u}_2,\d)}\right),\\
&&\\
x&=&\left[-12t\dfrac{P^{g+1}_1(r,\d)}{P^{g+1}_0(r,\d)}+
\dfrac{R^{g+1}(r,\d)}{P^{g+1}_0(r,\d)}\right]_{r=\tilde{u}_2},\\
&&\\
x&=&\left[-12t\dfrac{P^{g+1}_1(r,\d)}{P^{g+1}_0(r,\d)}+
\dfrac{R^{g+1}(r,\d)}{P^{g+1}_0(r,\d)}\right]_{r=u_i},\quad
i=1,\dots,2g+1\\
\end{array}\right.
\eeqa
The reason to write the hodograph transform in the form (\ref{ug1}) is that
system (\ref{ug1}) is non degenerate at the phase transition 
$\tilde{u}_1=\tilde{u}_2=v$ when $v\in (u_{2j},u_{2j-1}),\;\;1\leq j\leq
g+1,\;\;u_{g+2}=a$. 

\noindent
From Theorem~\ref{theoT} and proposition~\ref{zerophi}
we deduce that  the last $2g+1$ equations in 
(\ref{ug1}) in the limit $\tilde{u}_1=\tilde{u}_2$ reduce to those of
system (\ref{T1}).

\noindent
Next we investigate the behavior of
the Abelian differentials of the second kind $\sg_k^{g+1}(r,\d,v)$, $k\geq 0$,  in the limit $\d\ra 0$, when $v$ belongs to one of the gaps (\ref{gap}). We have that  \cite{Fay}
\begin{equation}
\label{dptrail}
\sg_k^{g+1}(r,\d,v)\simeq \sg_k^g(r)-\dfrac{1}{\log\d}\,\om^g_v(r)
\int_{Q^-(v)}^{Q^+(v)}\sg_k^g(\xi),
\end{equation}
where $\om^g_v(r)$  has been defined in (\ref{CK})
 and $Q^{\pm}(v)=(v,\pm\mu(v))$.  
From (\ref{dptrail}) we can obtain the expansion for the
polynomial $P^{g+1}_k(r,\d,v)=\tilde{\mu}(r)\dfrac{\sg_k^{g+1}(r,\d,v)}{dr}$,
namely
\begin{equation}
\label{dptrailp}
P_k^{g+1}(r,\d,v)\simeq (r-v) P_k^g(r)-\dfrac{1}{\log\d}\,(r-v)\mu(r)\om^g_v(r)
\int_{Q^-(v)}^{Q^+(v)}\sg_k^g(\xi),
\end{equation}
so that 
\begin{equation}
\label{ptrail}
\left.\dfrac{P_k^{g+1}(r,\d)}{P_0^{g+1}(r,\d)}\right|_{r=v\pm\sqrt\d}\simeq
\dfrac{\int_{Q^-(v)}^{Q^+(v)}\sg_k^g(\xi)}{\int_{Q^-(v)}^{Q^+(v)}\sg_0^g(\xi)}\left(1
\mp\sqrt{\d}\log\d\left(\dfrac{\sg_k^g(v)}{\int_{Q^-(v)}^{Q^+(v)}\sg_k^g(\xi)}-
\dfrac{\sg_0^g(v)}{\int_{Q^-(v)}^{Q^+(v)}\sg_0^g(\xi)}\right)\right)
\end{equation}
\noindent

Using (\ref{R2g}), (\ref{z1}-\ref{z2}) and (\ref{ptrail}) we have that 

\begin{equation}
\label{temp}
\begin{split}
\left.\dfrac{R^{g+1}(\tilde{u}_i,\d)}{P^{g+1}_0(\tilde{u}_i,\d)}\right|_{\tilde{u}_1=
\tilde{u}_2}=& \sum_{k=1}^{g}q^g_k(\vu)
\sum_{l=1}^k(2l-1)
\tilde{\Gamma}^{g}_{k-l}(\vu)\dfrac{\int_{Q^-(v)}^{Q^+(v)}\sg_k^g(\xi)}{\int_{Q^-(v)}^{Q^+(v)}\sg_0^g(\xi)}\\
+&q_{g+1}^{g+1}(v,v,\vu)\sum_{l=1}^{g+1}(2l-1)
\tilde{\Gamma}^{g}_{g+1-l}(\vu)\dfrac{\int_{Q^-(v)}^{Q^+(v)}\sg_k^g(\xi)}{\int_{Q^-(v)}^{Q^+(v)}\sg_0^g(\xi)},\; i=1,2.
\end{split}
\end{equation}
Adding and subtracting 
\[
\dfrac{2}{\int_{Q^-(v)}^{Q^+(v)}\sg_0^g(\xi)}\sum_{i=1}^{2g+1}
\int_{Q^-(v)}^{Q^+(v)}\dfrac{\mu(\xi)d\xi}{\xi-u_i}
\partial_{u_i}q_g(\vu),
\]
 to (\ref{temp}) and using (\ref{mu}) and (\ref{psir}) we obtain
\begin{equation}
\label{d1}
\left.\dfrac{R^{g+1}(\tilde{u}_i,\d)}{P^{g+1}_0(\tilde{u}_i,\d)}\right|_{\tilde{u}_1=
\tilde{u}_2}=\dfrac{1}{\int_{Q^-(v)}^{Q^+(v)}
\sg_0^g(\xi)}
\int_{Q^-(v)}^{Q^+(v)}\dfrac{2\mu^2(\xi)\Phi^g(\xi;\vu)+R^{g}(\xi)}{\mu(\xi)}d\xi, \;\;i=1,2
\end{equation}
where $\Phi^g(r;\vu)$ and $R^{g}(r)$ have been defined in (\ref{Phi}) and
(\ref{R2g}) respectively.
We observe that 
\[
\int_{Q^-(v)}^{Q^+(v)}\sg_0^g(\xi)\neq 0
\]
for all $v$ belonging to the gaps (\ref{gap}).

We need to  consider also the quantity in the first equation of
(\ref{ug1}) namely
\begin{equation}
\nonumber
\begin{split}
&\left.\left(\dfrac{-1}{\log(\frac{\tilde{u}_1-\tilde{u}_2}{2})^2(\tilde{u}_1-\tilde{u}_2)}
\left(\dfrac{R^{g+1}(\tilde{u}_1,\d)}{P^{g+1}_0(\tilde{u}_1,\d)}-
\dfrac{R^{g+1}(\tilde{u}_2,\d)}{P^{g+1}_0(\tilde{u}_2,\d)}\right)\right)\right|_{\tilde{u}_1=\tilde{u}_2}=\\
-&8\left.\left(\dfrac{\sqrt\d}{\log\d}\dfrac{\mu^2(v+\sqrt\d)}{P_0^{g+1}(v+\sqrt\d,\d)}\partial_{\tilde{u}_1}\partial_{\tilde{u}_2}q^{g+1}_{g+1}(\tilde{u}_1,\tilde{u}_2,\vu)\right)\right|_{\d=0}\\
-&2\left.\left(\left(\dfrac{\mu^2(v+\sqrt\d)}{\log\d\,P_0^{g+1}(v+\sqrt\d,\d)}+\dfrac{\mu^2(v-\sqrt\d)}{\log\d\,P_0^{g+1}(v-\sqrt\d,\d)}\right)\partial_{\tilde{u}_2}q^{g+1}_{g+1}(\tilde{u}_1,\tilde{u}_2,\vu)\right)\right|_{\d=0}\\
-&\sum_{k=1}^{g}q^g_k(\vu)
\sum_{l=1}^k(2l-1)
\tilde{\Gamma}^{g}_{k-l}(\vu)\left.\left(\dfrac{1}{2\sqrt\d\log\d}\left(\dfrac{P^{g+1}_{l-1}(v+\sqrt\d,\d)}{ P^{g+1}_0(v+\sqrt\d,\d)}-\dfrac{P^{g+1}_{l-1}(v-\sqrt\d,\d)}{ P^{g+1}_0(v-\sqrt\d,\d)}\right)\right)
\right|_{\d=0}\\
-& q_{g+1}^{g+1}(v,v,\vu)\sum_{l=1}^{g+1}(2l-1)
\tilde{\Gamma}^{g}_{g+1-l}(\vu)\left.\left(\dfrac{1}{2\sqrt\d\log\d}\left(\dfrac{P^{g+1}_{l-1}(v+\sqrt\d,\d)}{P^{g+1}_0(v+\sqrt\d,\d)}-\dfrac{P^{g+1}_{l-1}(v-\sqrt\d,\d)}
{ P^{g+1}_0(v-\sqrt\d,\d)}\right)\right)
\right|_{\d=0}
\end{split}
\end{equation}
Using (\ref{mu}), 
(\ref{psir}), (\ref{ptrail}) and (\ref{qpsi}) we simplify the above relation to the
form
\begin{equation}
\nonumber
\begin{split}
&\left.\left(\dfrac{-1}{\log(\frac{\tilde{u}_1-\tilde{u}_2}{2})^2(\tilde{u}_1-\tilde{u}_2)}
\left(\dfrac{R^{g+1}(\tilde{u}_1,\d)}{P^{g+1}_0(\tilde{u}_1,\d)}-
\dfrac{R^{g+1}(\tilde{u}_2,\d)}{P^{g+1}_0(\tilde{u}_2,\d)}\right)\right)\right|_{\tilde{u}_1=\tilde{u}_2}=\dfrac{1}{\mu(v)\int_{Q^-(v)}^{Q^+(v)}\sg_0^g(\xi)}\times\\
&\left(  2\mu^2(v)\partial_v\Psi^g(v;\vu)
+\sum_{k=1}^{g}q^g_k(\vu)
\sum_{l=1}^k(2l-1)
\tilde{\Gamma}^{g}_{k-l}(\vu)
\left(   P_{l-1}^g(v)-P_0^g(v)\dfrac{\int_{Q^-(v)}^{Q^+(v)}\sg_{l-1}^g(\xi)} 
{\int_{Q^-(v)}^{Q^+(v)}\sg_0^g(\xi)}\right)\right.\\
+&\left.\left(2\mu(v)\mu^{\prime}(v)-
4\mu(v)\dfrac{P_0^g(v,\vu)}{\int_{Q^-(v)}^{Q^+(v)}\sg_0^g(\xi)}
\right)\Psi^g(v;\vu)\right)
\end{split}
\end{equation}
Using repeatedly (\ref{psir}) we can write the above relation in the form
\begin{equation}
\label{d2}
\begin{split}
&\left.\left(\dfrac{-1}{\log(\frac{\tilde{u}_1-\tilde{u}_2}{2})^2(\tilde{u}_1-\tilde{u}_2)}
\left(\dfrac{R^{g+1}(\tilde{u}_1,\d)}{P^{g+1}_0(\tilde{u}_1,\d)}-
\dfrac{R^{g+1}(\tilde{u}_2,\d)}{P^{g+1}_0(\tilde{u}_2,\d)}\right)\right)\right|_{\tilde{u}_1=\tilde{u}_2}=\dfrac{1}{\mu(v)\int_{Q^-(v)}^{Q^+(v)}\sg_0^g(\xi)}\times\\
&\left(
2\mu^2(v)\Phi^g(v;\vu)+R^{g}(v)-\dfrac{P_0^g(v)}{\int_{Q^-(v)}^{Q^+(v)}\sg_0^g(\xi)}\int_{Q^-(v)}^{Q^+(v)}\dfrac{2\mu^2(\xi)\Phi^g(\xi;\vu)+
R^{g}(\xi)}{\mu(\xi)}d\xi\right),
\end{split}
\end{equation}
where $\Phi^g(r;\vu)$ and $R^{g}(r)$ have been defined in (\ref{Phi}) and
(\ref{R2g}) respectively.

\noindent
From  (\ref{ptrail}), (\ref{d1}) and (\ref{d2}) 
   system (\ref{ug1}) can be reduced  to the  form
\beqa
\label{ug2}
\left\{\begin{array}{lll}
0= 2\mu^2(v)\Phi^g(v;\vu)+Z^g(v)+
\dfrac{P_0^g(v)}{\int_{Q^-(v)}^{Q^+(v)}\sg_0^g(\xi)}
\displaystyle\int_{Q^-(v)}^{Q^+(v)}\;\dfrac{2\mu^2(\xi)\Phi^g(\xi;\vu)+Z^g(\xi)}{\mu(\xi)}d\xi&&\\
&&\\
0=\displaystyle\int_{Q^-(v)}^{Q^+(v)}\;\dfrac{2\mu^2(\xi)\Phi^g(\xi;\vu)+Z^g(\xi)}{\mu(\xi)}d\xi&&\\
&&\\
x=\left[-t\dfrac{P^{g}_1(r)}{P^{g}_0(r)}+
\dfrac{R^{g}(r)}{P^{g}_0(r)}\right]_{r=u_i},\quad
i=1,\dots,2g+1&&\\
\end{array}\right.
\eeqa
where the polynomial $Z^g(r)$ has been defined in (\ref{zeta}).
From proposition~\ref{zerop} the last $2g+1$ equations in (\ref{ug2})
are equivalent to the condition
\[
Z^g(r)\equiv 0\,, \;\;g>0.
\]
Therefore  system (\ref{ug2}) is equivalent for $g>0$  to the system 
\beqa
\label{ug3}
\left\{\begin{array}{lll}
0=\Phi^g(v;\vu)&&\\
&&\\
0=\displaystyle\int_{Q^-(v)}^{Q^+(v)}\;\mu(\xi)\Phi^g(\xi;\vu) d\xi&&\\
&&\\
x=\left[-t\dfrac{P^{g}_1(r)}{P^{g}_0(r)}+
\dfrac{R^{g}(r)}{P^{g}_0(r)}\right]_{r=u_i},\quad
i=1,\dots,2g+1.\;&&\\
\end{array}\right.
\eeqa
Because of  (\ref{normphi}),  when  $v\in (u_{2j},u_{2j-1}), 
1\leq j\leq g+1,\;u_{2g+2}=a$,  
  we can split the 
 integral of the second equation of (\ref{ug3}) in the form 
\begin{equation}
\nonumber
\begin{split}
\int_{Q^-(v)}^{Q^+(v)}\;\mu(\xi)\Phi^g(\xi;\vu) d\xi=& 
\int_{Q^-(v)}^{u_{2j-1}}\;\mu(\xi)\Phi^g(\xi;\vu) d\xi+\int_{u_{2j-1}}^{Q^+(v)}\;\mu(\xi)\Phi^g(\xi;\vu) d\xi\\
=&2\int_{u_{2j-1}}^{Q^+(v)}\;\mu(\xi)\Phi^g(\xi;\vu) d\xi.
\end{split}
\end{equation}
Therefore (\ref{ug3}) is equivalent to (\ref{T1}) for $g>0$.

\noindent
For $g=0$ substituting (\ref{zerodiff}) in 
(\ref{ug2}) it is easy to check that  we obtain a system equivalent to  (\ref{T1}).
\hfill$\square$ 
\vskip .2cm
\noindent

\vskip 0.5cm

\noindent
{\bf Proof of Theorem~\ref{CC}}.

\noindent
The equations describing  the   point
of gradient catastrophe of the $g$-phase solution 
can  be obtained either  considering the limit of the
$(g+1)-$phase solution when three   Riemann invariants coalesce, 
or supposing that one
of the $2g+1$ distinct Riemann invariants of the $g$-phase solution has
a vertical inflection point for $t>0$. 
For  proving Theorem~\ref{CC} we follow the latter possibility.

\noindent
On the solution of (\ref{ch}) 
 $\partial_x u_l(x,t)=(\partial_{u_l}(-\lb_lt+w_l))^{-1}$ \cite{FRT1} therefore
 a point of gradient catastrophe of the $g$-phase solution  
is determined by the system
\beqa
\label{cc1}
\left\{\begin{array}{lll}
\partial_{u_l}(-\lb_lt+w_l)=0&&\\
&&\\
(\partial_{u_l})^2(-\lb_lt+w_l)=0&&\\
&&\\
x=\left[-t\dfrac{P^{g}_1(r)}{P^{g}_0(r)}+
\dfrac{R^{g}(r)}{P^{g}_0(r)}\right]_{r=u_i},\quad
i=1,\dots,2g+1\\
\end{array}\right.
\eeqa 
where $1\leq l\leq 2g+1$.
We show that system (\ref{cc1}) is equivalent to system (\ref{CCe}).
For the purpose we compute explicitly the derivative $\partial_{u_l}(-\lb_lt+w_l)$.
From a generalization of a result in \cite{L} we obtain:
\begin{equation}
\label{partiall}
\dfrac{\partial}{\partial u_l}\dfrac{P^g_k(u_l)}{P^g_0(u_l)}=\dfrac{1}{2}
\left.\dfrac{\partial}{\partial r}\dfrac{P^g_k(r)}{P^g_0(r)}\right|_{r=u_l},
\end{equation}
so that 
\begin{equation}
\label{asd}
\begin{split}
&\dfrac{\partial}{\partial u_l}(-\lb_l(\vu)t+w_l(\vu))= -6t
\dfrac{\partial}{\partial r} \left.\dfrac{P^g_1(r)}{P_0^g(r)} 
\right|_{r=u_l}
+\dfrac{ \sum_{\overset{k=1}{k\neq l}}^{2g+1}
\prod_{\overset{m=1}{m\neq k,l}}^{2g+1}(u_l-u_m) }{P^g_0(u_l)} 
\partial_{u_l}q_g(\vu) \\
+& \dfrac{\prod_{\overset{m=1}{m\neq l}}^{2g+1}(u_l-u_m) }
{P^g_0(u_l)}(\partial_{u_l})^2
q_g(\vu)-\dfrac{\prod_{\overset{m=1}{m\neq l}}^{2g+1}(u_l-u_m) }
{(P^g_0(u_l))^2}
\partial_{u_l}q_g(\vu)\partial_{u_l}P^g_0(u_l)\\
+&\dfrac{1}{2}\sum_{n=1}^g (2n-1) \dfrac{\partial}{\partial r}\left.
\dfrac{P^g_{l-1}(r)}{P_0^g(r)}\right|_{r=u_l}\sum_{m=n}^g q_m(\vu)\Gt_{m-n}
+\sum_{n=1}^g (2n-1)\dfrac{P^g_{n-1}(u_l)}{P_0^g(u_l)}\sum_{m=n}^g 
\partial_{u_l}(q_m(\vu)\Gt_{m-n}).\\
\end{split}
\end{equation}
In the above relation we need to compute the derivative 
\[
\partial_{u_l}P^g_0(u_l)=\partial_r P^g_0(r)|_{r=u_l}+
\partial_{u_l} P^g_0(r)|_{r=u_l},
\]
where $P^g_0(r)=r^g+\alpha_1^0r^{g-1}+\dots+\alpha_g^0$. 
 For computing the derivatives of the
 normalization constants $\alpha_1^0,\alpha_2^0,\dots,\alpha_g^0$ 
in  $P^g_0(r)$ we need the following proposition.
\begin{prop}\cite{D1}
\label{dubrovin}
Let $\om_1(r)$ and $\om_2(r)$ two normalized Abelian differentials on
$\cs_g$. Let be $\xi=\dfrac{1}{\sqrt r}$ the
local coordinate at infinity and
\[
\om_1=\sum_k a^1_k \xi^k d\xi,\quad \om_2=\sum_k a^2_k\xi^k d\xi.
\]
Define the bilinear product
\[
V_{\om_1\om_2}=\sum_{k\geq 0}\dfrac{a^1_{-k-2}a^2_k}{k+1},
\]
then
\begin{equation}
\dfrac{\partial}{\partial u_i}V_{\om_1\om_2}=\res[r=u_i]\dfrac{\om_1(r)\om_2(r)}{dr},\quad i=1,\dots, 2g+1,
\end{equation}
where $\res[r=u_i]\dfrac{\om_1(r)\om_2(r)}{dr}$ is the residue of the differential
$\dfrac{\om_1(r)\om_2(r)}{dr}$ evaluated at $r=u_i$.
\end{prop}
Applying the above proposition to $\sg_0$ and $\sg_k$, $k=0,\dots g-1$
we obtain after non trivial simplifications
\begin{equation}
\begin{split}
\dfrac{\partial}{\partial u_l}
\begin{pmatrix}
\alpha_1^0\\
\alpha_2^0\\
\dots\\
\alpha_{g-1}^0\\
\alpha_g^0
\end{pmatrix}=&
-\dfrac{1}{2}
\begin{pmatrix}
1\\
u_l\\
\dots\\
u_l^{g-1}\\
u_l^{g-1}
\end{pmatrix}
-\dfrac{1}{2}
\begin{pmatrix}
0&0&0&\dots&0\\
1&0&0&\dots&0\\
\dots&\dots&\dots&\dots&\dots\\
u_l^{g-3}&u_l^{g-4}&\dots&0&0\\
u_l^{g-2}&u_l^{g-3}&\dots&1&0
\end{pmatrix}
\begin{pmatrix}
\alpha_1^0\\
\alpha_2^0\\
\dots\\
\alpha_{g-1}^0\\
\alpha_g^0
\end{pmatrix}\\+&
\dfrac{1}{2}\dfrac{P^g_0(u_l)}{\prod_{k=1,k\neq l}^{2g+1}(u_l-u_k)}
\begin{pmatrix}
\Gt_0&0&0&\dots&0\\
\Gt_1&\Gt_0&0&\dots&0\\
\dots&\dots&\dots&\dots&\dots\\
\Gt_{g-2}&\Gt_{g-3}&\dots&\Gt_0&0\\
\G_{g-1}&\Gt_{g-2}&\dots&\Gt_1&\Gt_0
\end{pmatrix}
\begin{pmatrix}
1\\
3P^g_1(u_l)\\
\dots\\
(2g-3)P^g_{g-2}(u_l)\\
(2g-1)P^g_{g-1}(u_l)
\end{pmatrix}
\end{split}
\end{equation}
where the $\Gt_k$'s have been defined in (\ref{Gt}).
From the above formula we obtain
\begin{equation}
\label{partialP}
\begin{split}
\partial_{u_l}P^g_0(u_l)=&\partial_r P^g_0(r)|_{r=u_l}+
\partial_{u_l} P^g_0(r)|_{r=u_l}\\
=&\dfrac{1}{2} \partial_r P^g_0(r)|_{r=u_l}+\dfrac{1}{2}
\dfrac{ P^g_0(u_l) }{ \prod_{k=1,k\neq l}^{2g+1}(u_l-u_k) }
\sum_{n=1}^g (2n-1)P^g_{n-1}(u_l)\sum_{m=n}^g u_l^m\Gt_{m-n}.
\end{split}
\end{equation}

Using  the relations (\ref{relations}), (\ref{dGt}),
(\ref{partiall}) and (\ref{partialP}) we simplify (\ref{asd}) to the form 
\begin{equation}
\label{AQ}
\begin{split}
\dfrac{\partial}{\partial u_l}(-\lb_l(\vu)t+&w_l(\vu))=
-\dfrac{ Z^g(u_l) }{2(P^g_0(u_l))^2}\partial_rP^g_0(r)|_{r=u_l}
+\dfrac{ \sum_{\overset{k=1}{k\neq l}}^{2g+1}
\prod_{\overset{m=1}{m\neq k,l}}^{2g+1}(u_l-u_m) }{P^g_0(u_l)}
\partial_{u_l}q_g(\vu) \\
+&\dfrac{1}{2P^g_0(u_l)} \partial_r\left.\left(\sum_{n=1}^g (2n-1)P^g_{n-1}(r)\sum_{m=n}^g q_m\Gt_{m-n}-xP^g_0(r)-12tP^g_1(r)\right)\right|_{r=u_l},\\
\end{split}
\end{equation}
where the polynomial $Z^g(r)$ has been defined in (\ref{zeta}).
Applying in the second term of (\ref{AQ}) the relations (\ref{psir}) 
 we can rewrite (\ref{AQ}) in the compact form
\begin{equation}
\label{aw1}
\dfrac{\partial}{\partial u_l}(-\lb_l(\vu)t+w_l(\vu))=
-\left.\dfrac{\partial}{\partial r}\dfrac{Z^g(r) }{ 2P^g_0(r) }\right|_{r=u_l}+
\dfrac{\prod_{\overset{m=1}{m\neq l}}^{2g+1}(u_l-u_m) }{P^g_0(u_l)}
\Phi^g(u_l;\vu),
\end{equation}
where $\Phi^g(r;\vu)$ has been  defined in  (\ref{Phi}). 
From proposition~\ref{zerop} the last $2g+1$ equations in (\ref{cc1})
are equivalent to the condition
\[
Z^g(r)\equiv 0\,, \;\;g>0.
\]
 Therefore we can simplify (\ref{aw1}) to the form
\begin{equation}
\label{dd1}
\dfrac{\partial}{\partial u_l}(-\lb_l(\vu)t+w_l(\vu))=
\dfrac{\prod_{\overset{m=1}{m\neq l}}^{2g+1}(u_l-u_m) }{P^g_0(u_l)}\Phi^g(u_l;\vu),
\end{equation}
when $u_1>u_2>\dots>u_{2g+1}$ satisfy  the $g$-phase Whitham equations.
As regarding the second derivative $\dfrac{\partial^2}{\partial u^2_l}(-\lb_l(\vu)t+w_l(\vu))$, from (\ref{dd1}) we obtain
\begin{equation}
\label{dd0}
\dfrac{\partial^2}{\partial u^2_l}(-\lb_l(\vu)t+w_l(\vu))=\dfrac{\partial}{\partial u_l}\left(\dfrac{\prod_{\overset{m=1}{m\neq l}}^{2g+1}(u_l-u_m) }{P^g_0(u_l)}\right)
\Phi^g(u_l;\vu)+\dfrac{\prod_{\overset{m=1}{m\neq l}}^{2g+1}(u_l-u_m) }{P^g_0(u_l)}\partial_{u_l}\Phi^g(u_l;\vu).
\end{equation}
Observing that 
\[
\partial_{u_l}\Phi^g(u_l;\vu)=\partial_{r}\Phi^g(r;\vu)|_{r=u_l}+\partial_{u_l}\Phi^g(r;\vu)|_{r=u_l}, \quad \partial_{r}\Phi^g(r;\vu)|_{r=u_l}=2\partial_{u_l}\Phi^g(r;\vu)|_{r=u_l},
\]
we obtain the relation
$\partial_{u_l}\Phi^g(u_l;\vu)=\dfrac{3}{2}\partial_{r}\Phi^g(r;\vu)|_{r=u_l}$.
Therefore 
\begin{equation}
\label{dd2}
\dfrac{\partial^2}{\partial u^2_l}(-\lb_l(\vu)t+w_l(\vu))=\dfrac{\partial}{\partial u_l}\left(\dfrac{\prod_{m=1,m\neq l}^{2g+1}(u_l-u_m) }{P^g_0(u_l)}\right)
\Phi^g(u_l;\vu)+\dfrac{3}{2}\dfrac{\prod_{\overset{m=1}{m\neq l}}^{2g+1}(u_l-u_m) }{P^g_0(u_l)}\partial_{r}\Phi^g(r;\vu)|_{r=u_l}.
\end{equation}
 From  (\ref{dd1}-\ref{dd2}) and the fact that $\dfrac{\prod_{m=1,m\neq l}^{2g+1}(u_l-u_m) }{P_0(u_l)}\neq 0$ for $u_1>u_2>\dots>u_{2g+1}$,
it is clear that system (\ref{cc1})
is equivalent to system (\ref{CCe}).
\hfill $\square$
\vskip .2cm
We remark that to avoid higher order degeneracies in system (\ref{CCe}), 
we impose the condition
\begin{equation}
\label{terzo}
(\partial_r)^2\Phi^g(r;\vu)_{r=u_l(x_c,t_c)}\neq 0. 
\end{equation}
Indeed we  prove that the above condition guarantees that the genus of 
solution of the Whitham equations increases at most by one in the
neighborhood of the point of gradient catastrophe. It is sufficient
to show that the transition  to genus $g+2$ does not occur. For the purpose 
let us suppose $l$ even and let us  consider the function
\begin{equation}
\label{secondd}
0\neq (\partial_r)^2\Phi^g(r;\vu)_{r=u_l(x_c,t_c)}=\mbox{const}\;\times 
\Phi^{g+2}(u_l;u_1,\dots, u_{l-1},
u_l,u_l,u_l,u_l,u_l,u_{l+1},\dots,u_{2g+1}).
\end{equation}
By proposition~\ref{zphi}
 in order to have a genus $g+2$ solution in the neighborhood of the
 point of gradient catastrophe,   the function
\[
\Phi^{g+2}(r;u_1,\dots, u_{l-1},
u_l+\e_1,u_l+\e_2,u_l+\e_3,u_l+\e_4,u_l+\e_5,u_{l+1},\dots,u_{2g+1})
\]
has to change sign in each of the intervals  $(u_l+\e_2,u_l+\e_3)$ 
and $(u_l+\e_4,u_l+\e_5)$
($l$ even) for arbitrary small $\e_1>\e_2>\e_3>\e_4>\e_5>0$. 
Because of (\ref{secondd})
\[
\Phi^{g+2}(r;u_1,\dots, u_{l-1},
u_l+\e_1,u_l+\e_2,u_l+\e_3,u_l+\e_4,u_l+\e_5,u_{l+1},\dots,u_{2g+1})\neq 0 
\]
for  sufficiently small $\e_1>\e_2>\e_3>\e_4>\e_5>0$ and 
for $r\in(u_l+\e_2,u_l+\e_5)$.
Therefore  the transition  to genus $g+2$ does not occur. 
We can exclude transitions from genus $g$ to genus $g+n$, 
$n>2$, by perturbations arguments.
Therefore it is
legitimate to consider the point of gradient catastrophe that solve (\ref{CCe})
and satisfies (\ref{terzo})  as a point of the boundary between the domains
$D_g$ and $D_{g+1}$. We remark that in the genus $g=0$ case, the  condition
(\ref{secondd}) is not essential as illustrated by theorem
(\ref{Tian})

\vskip 0.3cm

\noindent 
{\bf Proof of Theorem (\ref{Tmultiple})}

\begin{figure}[htb]
\centering
\mbox{\subfigure{\epsfig{figure=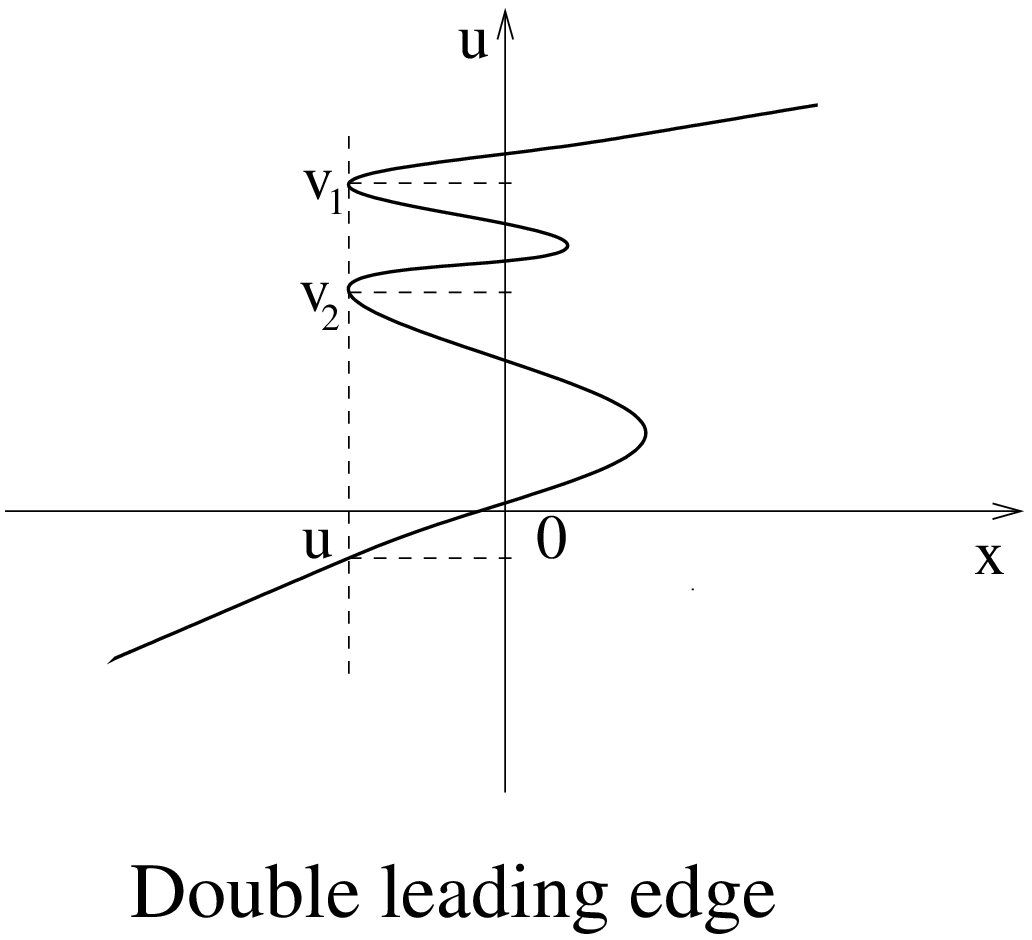,width=.3\textwidth}}
      \subfigure{\epsfig{figure=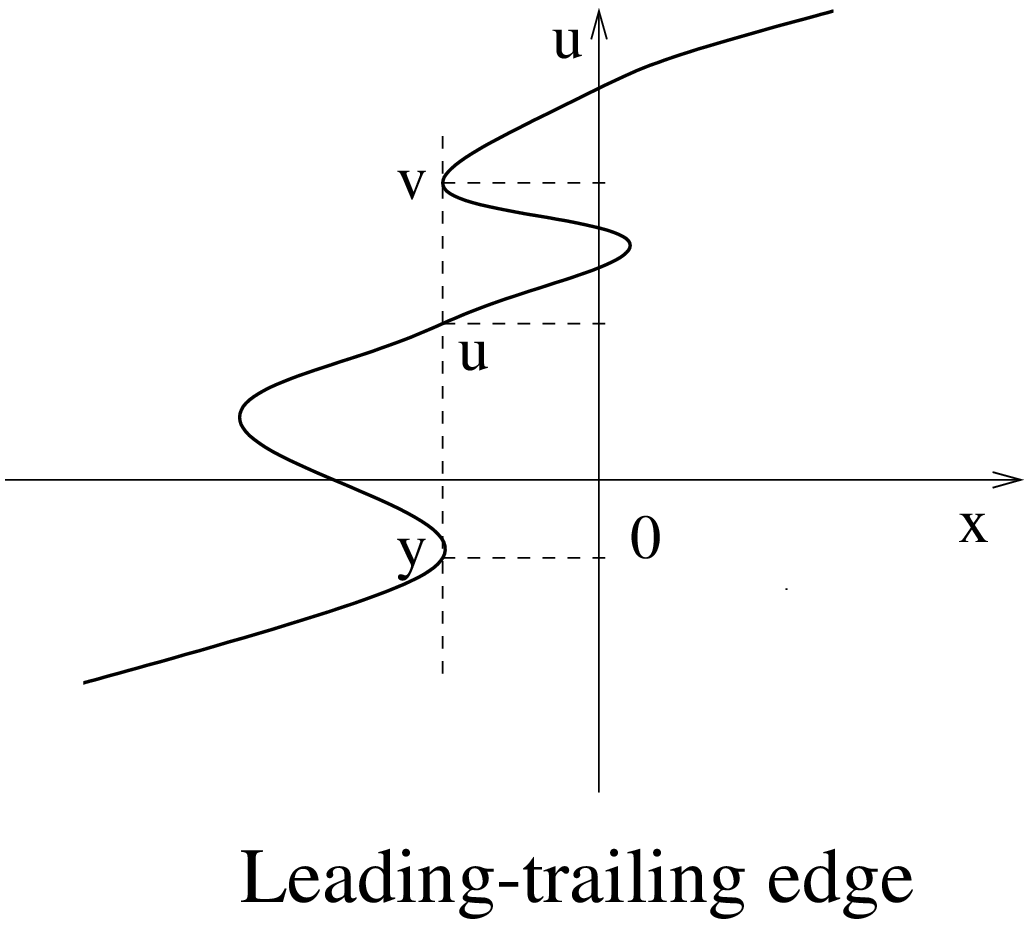,width=.3\textwidth}}
      \subfigure{\epsfig{figure=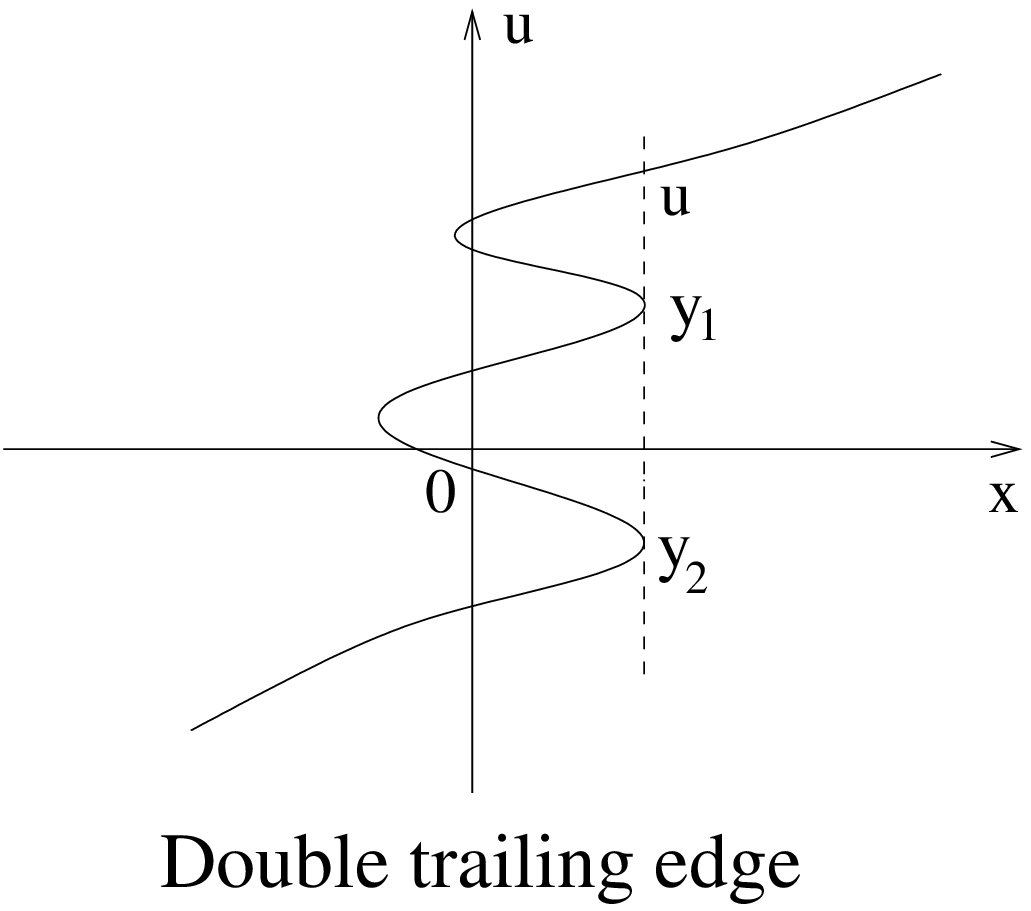,width=.3\textwidth}}}
\caption{Three kinds of phase transition from the zero-phase
solution to the two-phase solution \label{figpt}}
\end{figure}

\noindent
A phase transition may  occur between the 
zero-phase solution and the two-phase
solution. As shown on Figure~\ref{figpt} we can have a 
double leading edge, 
a trailing-leading edge and a double trailing edge. There are
also other types of boundary between the 
zero-phase solution and the two-phase
solution which we call  ``point of gradient catastrophe $\&$
leading edge'' and ``point of gradient catastrophe $\&$ trailing 
edge''.
Multiple  transitions may also occur  between the 
$g$-phase solution and the
$(g+2)$-phase solution. In order to determine the systems which describe
such phase transitions we  consider the Riemann surface $\cs_{g+2}$ of
genus $ g+2$ defined by 
\[
\tilde{\mu}^2=(r-v_1-\sqrt\d_1 )(r-v_1+\sqrt\d_1)
(r-v_2-\sqrt\d_2 )(r-v_2+\sqrt\d_2)\mu^2,\quad v\in\Rn\,,\;\;
\]
\[
\mu^2(r)=
\prod_{j=1}^{2g+1}(r-u_j),\quad
u_1>u_2>\dots>u_{2g+1},
\] 
where $v_j\neq u_i$, $j=1,2$, $i=1,\dots,2g+1$. Then  we study the
hodograph transform for the distinct variables 
$v_1\pm\sqrt\d_1$, $v_2\pm\sqrt\d_2$
and $u_1,\dots,u_{2g+1}$ in the {\it independent}  limits $\d_1\ra 0$ and $\d_2\ra 0$.
When $v_1$ and $v_2$ belong to the bands (\ref{band}), we are considering the double
leading edge. Repeating the calculations done  for the single leading edge 
we can determine the 
 equations which describe the phase transition for the double leading
edge, namely
\beqa
\label{doubleleadeq}
\left\{\begin{array}{lll}
\partial_{v_1}\Phi^g(v_1;\vu)=0&&\\
\Phi^g(v_1;\vu)=0&&\\
\partial_{v_2}\Phi^g(v_2;\vu)=0&&\\
\Phi^g(v_2;\vu)=0&&\\
x=\left[-12t\dfrac{P^{g}_1(r)}{P^{g}_0(r)}+
\dfrac{R^{g}(r)}{P^{g}_0(r)}\right]_{r=u_i}&&,\quad
i=1,\dots,2g+1,\; g>0\\
\end{array}\right.
\eeqa

From (\ref{y1}-\ref{T1}) analogous  systems
 can be obtained for the trailing-leading edge and
double trailing edge.

\noindent
For studying the phase 
 transitions  between the $g$-phase
solution and the $(g+n)$-phase solution, $n\geq 1$, having $n_1$
leading edges and $n_2$ trailing edges, $n_1+n_2=n$ we consider the
Riemann surface
\[
\tilde{\mu}^2=\prod_{j=1}^{n_1}(r-v_j-\sqrt\d_j )(r-v_j+\sqrt\d_j)\prod_{k=1}^{n_2}(r-y_k-\sqrt\rho_k )(r-y_k+\sqrt\rho_k)\mu^2,\;
\]
\[
\mu^2(r)=
\prod_{j=1}^{2g+1}(r-u_j),\quad
u_1>u_2>\dots>u_{2g+1}.
\] 
Here $v_j$, $j=1,\dots,n_1$, belongs to the bands (\ref{band}) and
$y_k$, $\;k=1\dots,n_2$, belongs to the gaps (\ref{gap}), 
$0<\d_j\ll 1$, $j=1,\dots,n_1$,
and $0<\rho_k\ll 1$, $\;k=1\dots,n_2$.

 We  study the
hodograph transform (\ref{ch}) for the variables $v_j\pm\sqrt\d_j$,
$j=1\dots,n_1$, $y_k\pm \sqrt\rho_k$, $\;k=1\dots,n_2$ 
and $u_1,\dots,u_{2g+1}$ in the {\it independent}  limits $\d_j\ra 0$,
$j=1\dots,n_1$
and $\rho_k\ra 0$ $\;k=1\dots,n_2$.
Repeating the calculations done for proving Theorem~\ref{LT} and
Theorem~\ref{TT}  it is easy to show that system (\ref{multiple})
describes the phase transition  between the $g$-phase
solution and the $(g+n)$-phase solution having $n_1$
leading edges, $n_2$ trailing edges and no points of gradient catastrophe, 
 $n_1+n_2=n>1$.  
If we suppose that the $g$-phase solution has also $n_3$ points of gradient
catastrophe, $n_1+n_2+n_3=n$, then combining  Theorem~\ref{LT},
Theorem~\ref{TT} and Theorem~\ref{CC} we obtain 
proposition~\ref{propom}.
\hfill $\square$
\vskip .2cm

\section{Conclusion}
In this work we have constructed,  in implicit form, 
the  $g$-phase solution of the
Whitham equations for monotone increasing smooth initial data $x=f(u)|_{t=0}$.
The goal is obtained solving  the Tsarev system
(\ref{tsarev}). 
We have shown that  the solution 
of the Tsarev system  which satisfies the natural  boundary conditions
 (\ref{b1}-\ref{b4}) is unique. 
Then we have investigated the conditions for the  solvability of the
hodograph transform (\ref{ch}). For the purpose
 we have derived all the equations
which describe a phase transition of the solution of the Whitham 
equations. Studying when  phase transitions occur we have
been able to prove the second main result of this work.
Namely we have shown   that when the initial data
satisfies (\ref{condition}), the solution  of the Whitham equations
has genus $g\leq N$ for all $x$ and $t\geq 0$.  It is still an open 
problem to effectively  determine, on the $x-t\geq 0$ plane, the function   
$0\leq g(x,t)\leq N$ from the generic initial data $x=f(u)|_{t=0}$. 
\vskip .5cm 

{\bf Acknowledgments.} I am indebted with Professor  Boris Dubrovin who 
posed me the problem of this work and gave me many hints 
to reach the solution. I am grateful to Professor Sergei Novikov for
his suggestions during the preparation of the manuscript.
This work was partially support by a CNR  grant 203.01.70 and by a grant of S. Novikov.


\end{document}